\documentclass[reprint, superscriptaddress, amsmath, amssymb, aps, pra, floatfix, longbibliography]{revtex4-1}
\usepackage{graphicx}
\usepackage{dcolumn}
\usepackage{bm}
\usepackage{textgreek}
\usepackage{color,soul}  
\usepackage{gensymb}
\usepackage{textcomp}
\usepackage{epsf}
\usepackage{relsize}
\usepackage[normalem]{ulem}
\usepackage{nicefrac}
\usepackage{comment}
\usepackage{hyperref}
\hypersetup{colorlinks,allcolors=black}
 \usepackage{natbib}
\usepackage{scalerel}
\usepackage{xcolor}
\usepackage{tikz}
\usetikzlibrary{svg.path}
\definecolor{orcidlogocol}{HTML}{A6CE39}
\tikzset{
  orcidlogo/.pic={
    \fill[orcidlogocol] svg{M256,128c0,70.7-57.3,128-128,128C57.3,256,0,198.7,0,128C0,57.3,57.3,0,128,0C198.7,0,256,57.3,256,128z};
    \fill[white] svg{M86.3,186.2H70.9V79.1h15.4v48.4V186.2z}
                 svg{M108.9,79.1h41.6c39.6,0,57,28.3,57,53.6c0,27.5-21.5,53.6-56.8,53.6h-41.8V79.1z M124.3,172.4h24.5c34.9,0,42.9-26.5,42.9-39.7c0-21.5-13.7-39.7-43.7-39.7h-23.7V172.4z}
                 svg{M88.7,56.8c0,5.5-4.5,10.1-10.1,10.1c-5.6,0-10.1-4.6-10.1-10.1c0-5.6,4.5-10.1,10.1-10.1C84.2,46.7,88.7,51.3,88.7,56.8z};
  }
}

\newcommand\orcid[1]{\href{https://orcid.org/#1}{\mbox{\scalerel*{
\begin{tikzpicture}[yscale=-1,transform shape]
\pic{orcidlogo};
\end{tikzpicture}
}{|}}}}

\usepackage{lineno}
\def \p {\partial}
\def \m {\mathbf }

\def\red{\textcolor{red}}

\def \be {\begin{equation}}
\def \ee {\end{equation}}
\def \bea {\begin{align}}
\def \eea {\end{align}}
\def \p {\partial}
\def\bee{\begin{eqnarray}}
\def\eee{\end{eqnarray}}
\def \BC {\begin{cases}}
\def \EC {\end{cases}}
\def \m {\mathbf }

\begin{document}

\title{Room Temperature  Amplification  of Terahertz Radiation by Grating-Gate Graphene Structures} 

\newcommand{\Sendai}{\affiliation{Research Institute of Electrical Communication, Tohoku University, Sendai 980-8577, Japan}}
\newcommand{\CENTERA}{\affiliation{CENTERA Laboratories, Institute of High Pressure Physics PAS, Warsaw 01-142, Poland}}
\newcommand{\CEZAMAT}{\affiliation{CEZAMAT Warsaw Technical University, Warsaw 02-346, Poland}}
\newcommand{\MLCC}{\affiliation{Laboratory Charles Coulomb, University of Montpellier and CNRS, Montpellier F-34095, France}}
\newcommand{\IRE}{\affiliation{Kotelnikov Institute of Radio Engineering and Electronics (Saratov Branch), RAS, Saratov 410019, Russia}}
\newcommand{\IOFFE}{\affiliation{ Ioffe  Institute, 194021 St. Petersburg, Russia}}

\author{Stephane Boubanga-Tombet}
\Sendai

\author{Wojciech Knap
\orcid{0000-0003-4537-8712}}
\Sendai
\CENTERA
\MLCC

\author{Deepika Yadav
\orcid{0000-0003-1240-9789}}
\Sendai

\author{Akira Satou
\orcid{0000-0002-4371-9344}}
\Sendai

\author{Dmytro~B.~But
\orcid{0000-0002-0735-4608}}
\CENTERA
\CEZAMAT

\author{Vyacheslav V. Popov
\orcid{0000-0003-1303-6443}}
\IRE

\author {Ilya V. Gorbenko}
\IOFFE

\author{Valentin Kachorovskii
\orcid{0000-0002-9684-2135}}
\CENTERA
\IOFFE

\author{Taiichi Otsuji
\orcid{0000-0002-0887-0479}}
\email{otsuji@riec.tohoku.ac.jp}
\Sendai

\date{\today} 

\begin{abstract}
 We study terahertz (THz) radiation transmission through grating-gate graphene based nanostructures. 
 We report on room temperature THz radiation amplification stimulated by current-driven plasmon excitation. 
 Specifically, with increase of the \textit{dc} current under periodic charge density modulation, we observe a strong  red shift of the resonant  THz  plasmon absorption, followed by a window of complete transparency to incoming radiation, and subsequent amplification and blue shift of the resonant plasmon frequency. 
 Our results are, to the best of our knowledge, the first experimental observation of  energy transfer from \textit{dc} current to plasmons leading to  THz amplification. Additionally, we present a simple model offering phenomenological description of the observed THz amplification. 
 This model shows that in the presence of \textit{dc} current the radiation-induced correction to dissipation is sensitive to the phase shift between oscillations of carrier density and drift velocity. 
 And with increasing current, the dissipation becomes negative, leading to amplification. 
 The experimental results of this work, as all obtained at room temperature, pave the way towards the new 2D plasmons based, voltage tuneable THz radiation amplifiers.
\end{abstract}



\maketitle


\section{Introduction}

More than forty years ago, active theoretical and experimental studies of plasma oscillations in two dimensional electron systems (2DESs) began and plasmonic resonances were observed~\cite{Chaplik1972, Allen1977a, Theis1977a, Tsui1978, Theis1978, Theis1980, Tsui1980a}. The interest to this area dramatically increased after seminal work of Dyakonov and Shur~\cite{Dyakonov1993a}, who theoretically predicted that the \textit{dc} current in the channel of a sub-micrometer size field effect transistors (FET) could become unstable leading to the excitation of plasma oscillations, with the frequency controlled by the gate voltage, and generation of tunable terahertz (THz) radiation.

This work, as well as the next publication~\cite{Dyakonov1996DetectionFluid} which was focused on the plasmon-mediated THz detection aroused great interest because of their novelty in the field of the fundamental physics and important potential applications in THz-optoelectronics for creation of all-electronic, compact, and gate-tunable THz detectors and  emitters. However, numerous experimental attempts to realize efficient, narrowband and voltage tunable  2D plasmon based detectors or emitters  of THz radiation with single FETs have failed as the intensity of radiation turned out to be too small, plasma resonances were too broad and/or not gate voltage tunable~\cite{Hirakawa1995, Knap2004TerahertzTransistors, Lusakowski2005VoltageTransistor, Dyakonova2005MagneticTransistors, Otsuji2008, Boubanga-Tombet2010, ElFatimy2010, Knap2011FieldEmission, Otsuji2013EmissionGraphene}. 

It did not take long to understand that multiple-gate periodic structures are more promising. Such structures interact much better with THz radiation than the single-gate structures.
In fact, study of the grating-gate-coupler-based plasmon excitation in two-dimensional electron gases by incident THz waves has begun a long time ago since the seminal works~\cite{Allen1977a, Theis1977a}. 
By using such structures, high-quality factor plasmon resonances in absorption could be excited, as it was demonstrated  much later  by Muravjov et al.~\cite{Muravjov2010TemperatureStructures}, for 2D grating gate GaN/AlGaN structures. The grating gate structures have also shown excellent characteristics as THz radiation broadband detectors~\cite{Kurita2014, Boubanga-Tombet2014, Faltermeier2015}. Nevertheless, despite tremendous efforts, neither the room temperature resonant detection, nor the current-stimulated emission or amplification of THz radiation have been observed so far (see discussion in the recent work \cite{Shalygin2019SelectiveHeterostructure}, where low temperature THz emission was discussed).

Fundamentally new aspects have been brought by the use of graphene-based structures~\cite{Neto2009, Bonaccorso2010}. Graphene shows record mobility at room temperature, which gives a much higher quality factor of plasma resonances than that in conventional materials. Moreover, the mediation of plasmons can greatly enhance the interaction between light and graphene substance due to relatively low level of losses, and a high degree of spatial electric field confinement. This explains the growing interest in graphene plasmonics~\cite{Vafek2006, Ryzhii2006TerahertzHeterostructures, Hwang2007DielectricGraphene, Ryzhii2007PlasmaHeterostructures, Jablan2009PlasmonicsFrequencies, West2010SearchingMaterials, Grigorenko2012GraphenePlasmonics, Koppens2011GrapheneInteractions}. 
Plasmon-enhanced THz graphene devices have recently been investigated~\cite{Bahk2014PlasmonGraphene, DeglInnocenti2016FastAntennas}, and improvement of the device performances on gain modulation~\cite{DeglInnocenti2016FastAntennas, Chakraborty2016AppliedLasers.}, sensitivity~\cite{Cai2015Plasmon-EnhancedGraphene} and emission~\cite{Bahk2014PlasmonGraphene} have already been demonstrated. 

Recently some theoretical and experimental studies of grating gate graphene structures utilizing plasma resonances have been reported~\cite{Bylinkin2019Tight-BindingGraphene, Svintsov2019EmissionTransport, Olbrich2015PRB}. In particular, the electromagnetic simulations of the graphene grating gate structures~\cite{Fateev2019PRA} have shown the possibility of the resonant plasmons strongly coupled to THz radiation even at room temperature.  
These works show that (i) the plasmon resonances can be excited
in graphene-based periodic   structures
in reflectivity configuration and (ii) such configuration can be used to obtain the THz radiation absorption by plasmons.

Despite intensive studies of the THz resonant detection in graphene-based periodic structures, there has not yet been any detailed study of current-induced effects. Also, there does not exist any report of room temperature experiments showing efficient energy transfer from \textit{dc} current to plasmons that could lead to THz radiation generation or amplification~\cite{Watanabe2013TheGraphene, Takatsuka2012GainStructures, Boubanga-Tombet2012UltrafastTemperature}.

In this work, we explore THz light-plasmon coupling, light absorption and amplification by graphene grating gate structures focusing on current-driven effects. These grating gate structures were used to create a periodic structure of highly-conducting active regions separated by low-conducting passive regions. The large difference between concentrations allowed us to localize plasmons in active regions and control their properties both by gate electrodes and by  the \textit{dc} driving current. We demonstrate that, in such structures, gate-voltage controlled resonant plasmons   are excited by THz radiation and  show current-driven amplification of this radiation. More specifically, we show that with the increase of \textit{dc} current the plasmon spectra undergoes a strong red shift, followed by complete lack of the resonant THz plasmon absorption, and subsequent amplification with a clear blue shift. 

We also present a phenomenological theoretical description of observed results using a simple model of periodically alternated stripes of the high/low electron density and corresponding high/low plasma wave velocity. We show that in such a plasmonic crystal structure, the THz radiation dissipation becomes sensitive to the phase shift between the oscillations of carrier density and drift velocity, and that, with increasing the \textit{dc} current, the radiation-related correction to dissipation changes its sign, resulting in amplification of the optical signal.

Our main results can be formulated as follows:

(i) Experiments demonstrating that the current-driven plasmon resonances at THz range, with the increase of the current, undergo red-shift, followed by a window of complete transparency to incoming radiation and subsequent amplification and blue-shift of the resonant plasmon frequency. 
(ii) Theory providing the phenomenological description of the experimentally observed phenomena, in particular, switching from dissipation to amplification at drift velocities smaller than the plasma wave velocity.

Importantly, all experimental results were obtained at room temperature, and therefore, can be used for design of graphene/semiconductors based resonant, compact and voltage/current controlled THz absorbers, amplifiers as well as coherent sources.

\section{\label{sec:method}Experimental}
\subsection{\label{sec:Sample}Samples}
\begin{figure}
\centering
(a)~~\includegraphics[width=0.2\textwidth,bb=0 0 118 120]{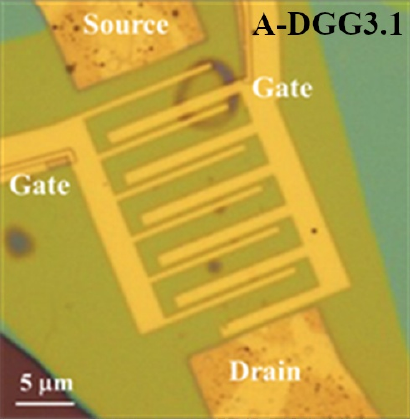}\par
(b)~~\includegraphics[width=0.7\linewidth]{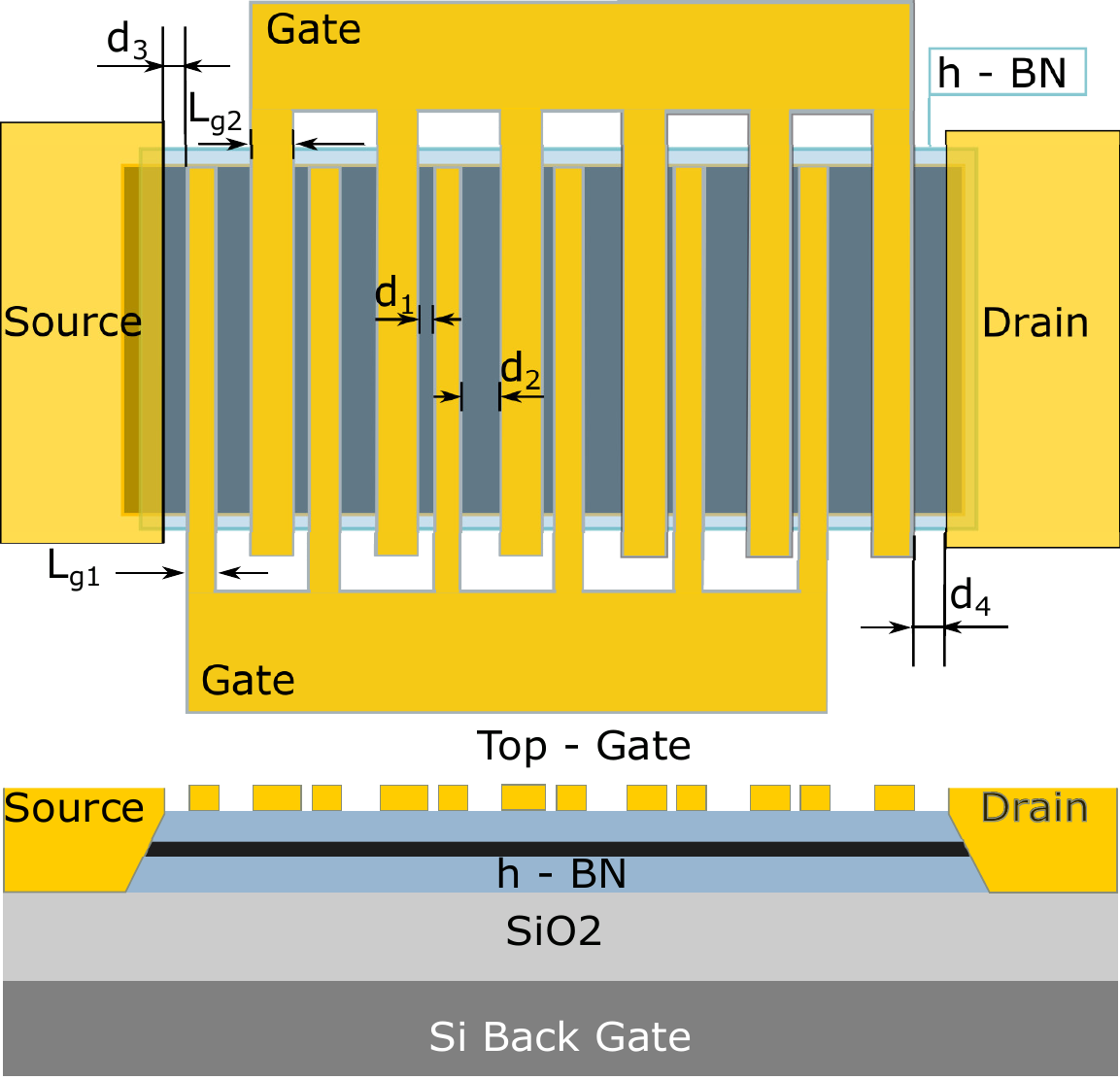}\par 
(c)~\includegraphics[width=0.94\linewidth]{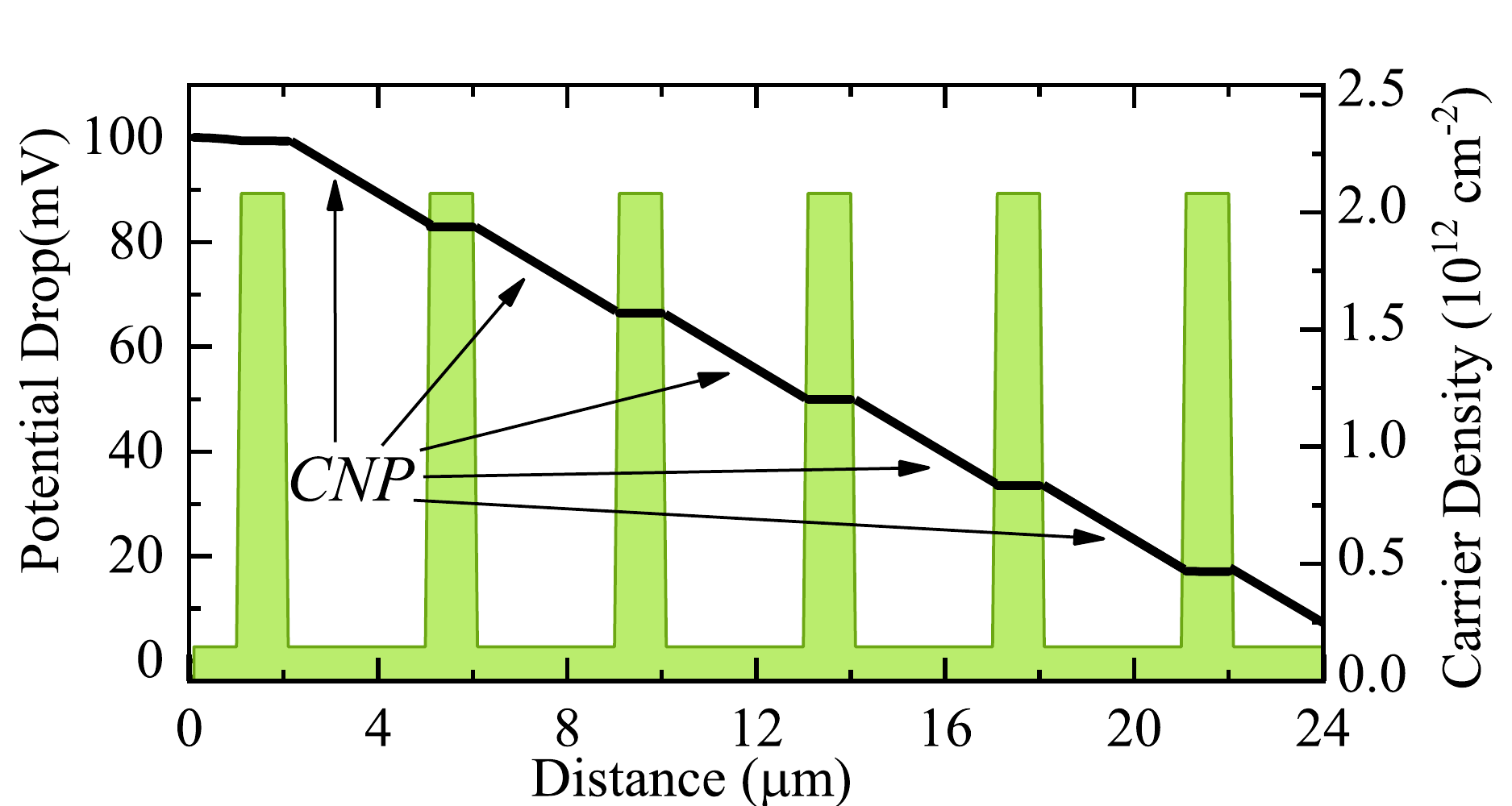}\par 
\caption{The microphoto of the A-DGG3.1 sample (a), schematic top and cross-sectional device images illustrating the h-BN/graphene/h-BN heterostructure including the asymmetric dual grating gate metallization (b), and the gate-controlled charge density distribution (the green area) together with potential distribution along the channel (the black solid line)(C).
}\label{Fig1}
\end{figure}

The samples were fabricated with a field-effect transistor structure featuring an interdigitated dual-grating-gate (DGG) where the plasmonic cavities are formed below the gates electrode grating fingers~\cite{Popov2011a, Boubanga-Tombet2014, Wilkinson1992a, Coquillat2010a}. In order to ensure high carrier mobility in our devices, h-BN-encapsulated graphene heterostructures (h-BN/graphene/h-BN) were fabricated. Optically estimated h-BN thickness was in the range from 20~nm to 32~nm for both structures. The first h-BN layer was transferred onto a SiO$_{2}$/Si wafer serving as a part of the back-gate dielectric layer. The process was followed by transferring the monolayer graphene. The second h-BN layer was transferred onto the monolayer graphene and served as the top gates dielectric layer. The layer sequence and the sample architecture are shown in Fig.~\ref{Fig1}. 
Raman spectroscopy (sharp mono peak at G and G$'$ band) and the contrast fractions of the optical microscope images of the transferred graphene sheets allowed confirming that they were monolayers. 

Two device structures were designed with gate fingers width of L$_{g1}$ = 0.5~\textmu m and L$_{g2}$ = 1~\textmu m (L$_{g1}$ = 0.75~\textmu m and L$_{g2}$ = 1.5~\textmu m) separated by d$_{1}$ = 0.5~\textmu m and d$_{2}$ = 1~\textmu m gaps (d$_{1}$ = 0.5~\textmu m and d$_{2}$ = 2~\textmu m) referred to as Asymmetric DGG  structures A-DGG 3.1 and (A-DGG 3.2), respectively (see Fig.~\ref{Fig1}). The devices had very similar electrical properties with very close Charge Neutrality Points (CNP), ranging from -0.1~V to +0.15~V. 
\begin{figure}[!ht]
\centering
(a)\includegraphics[width=0.92\linewidth]{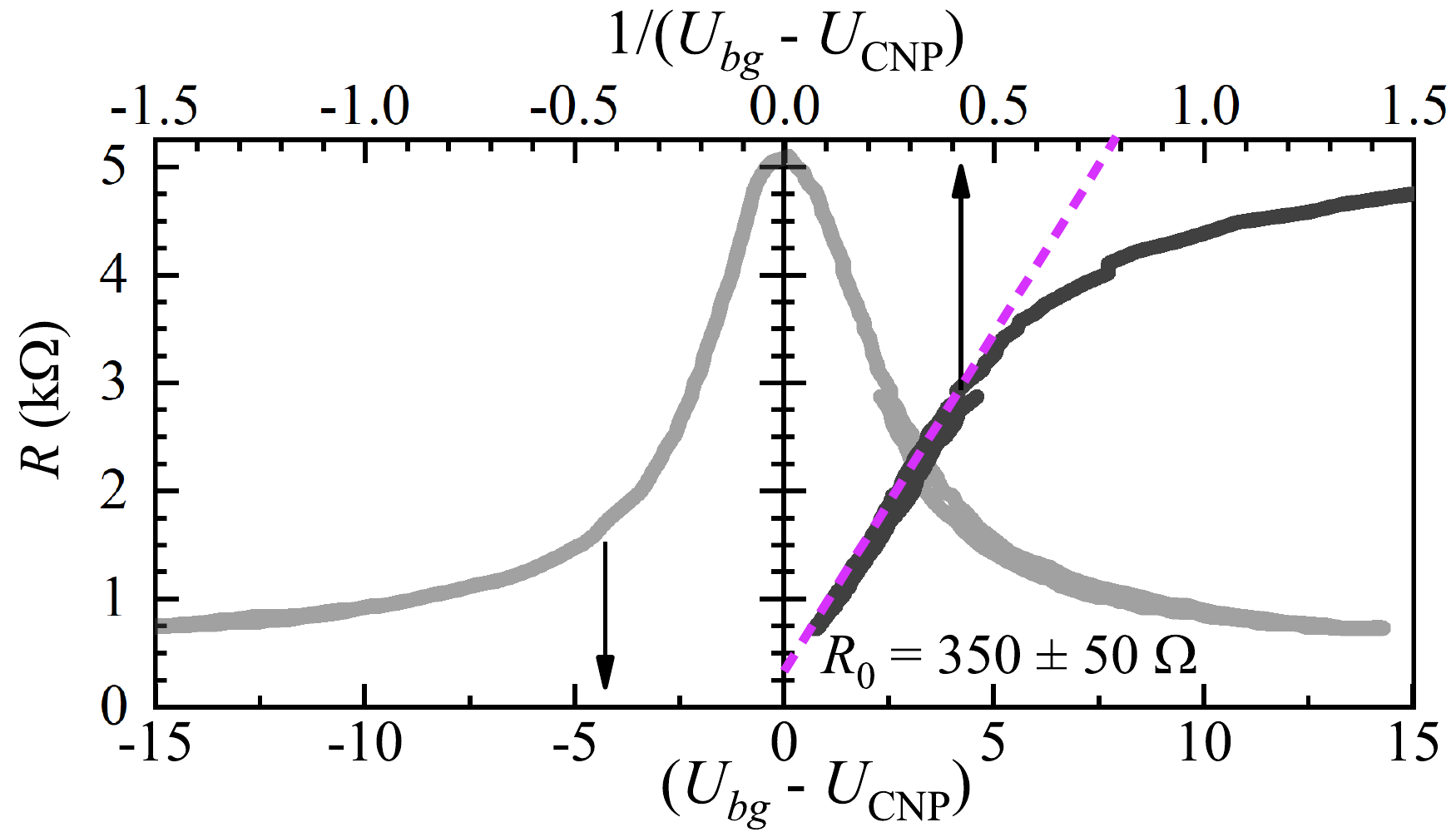}\par 
(b)\includegraphics[width=0.92\linewidth]{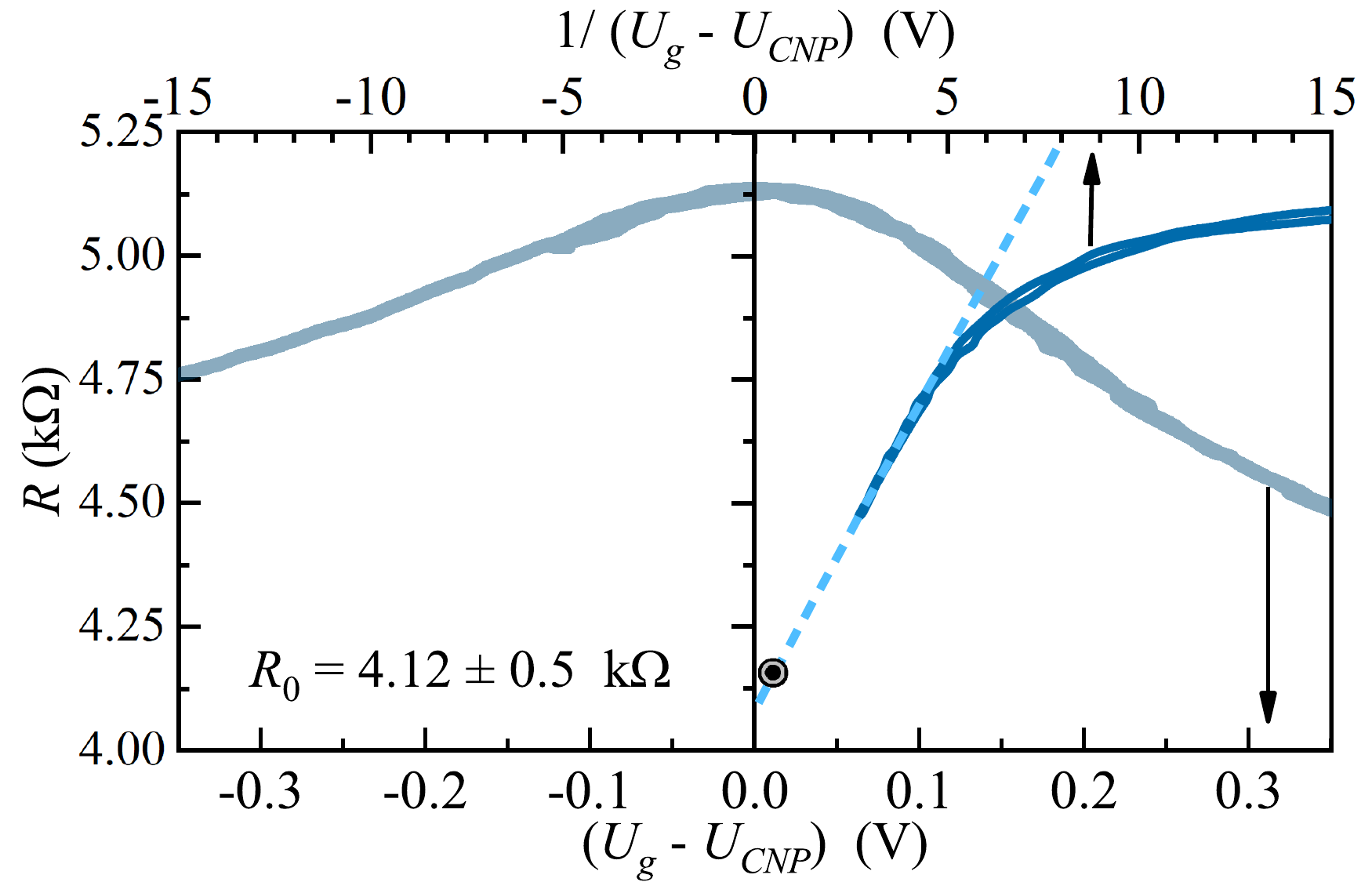}\par 
(c)\includegraphics[width=0.92\linewidth]{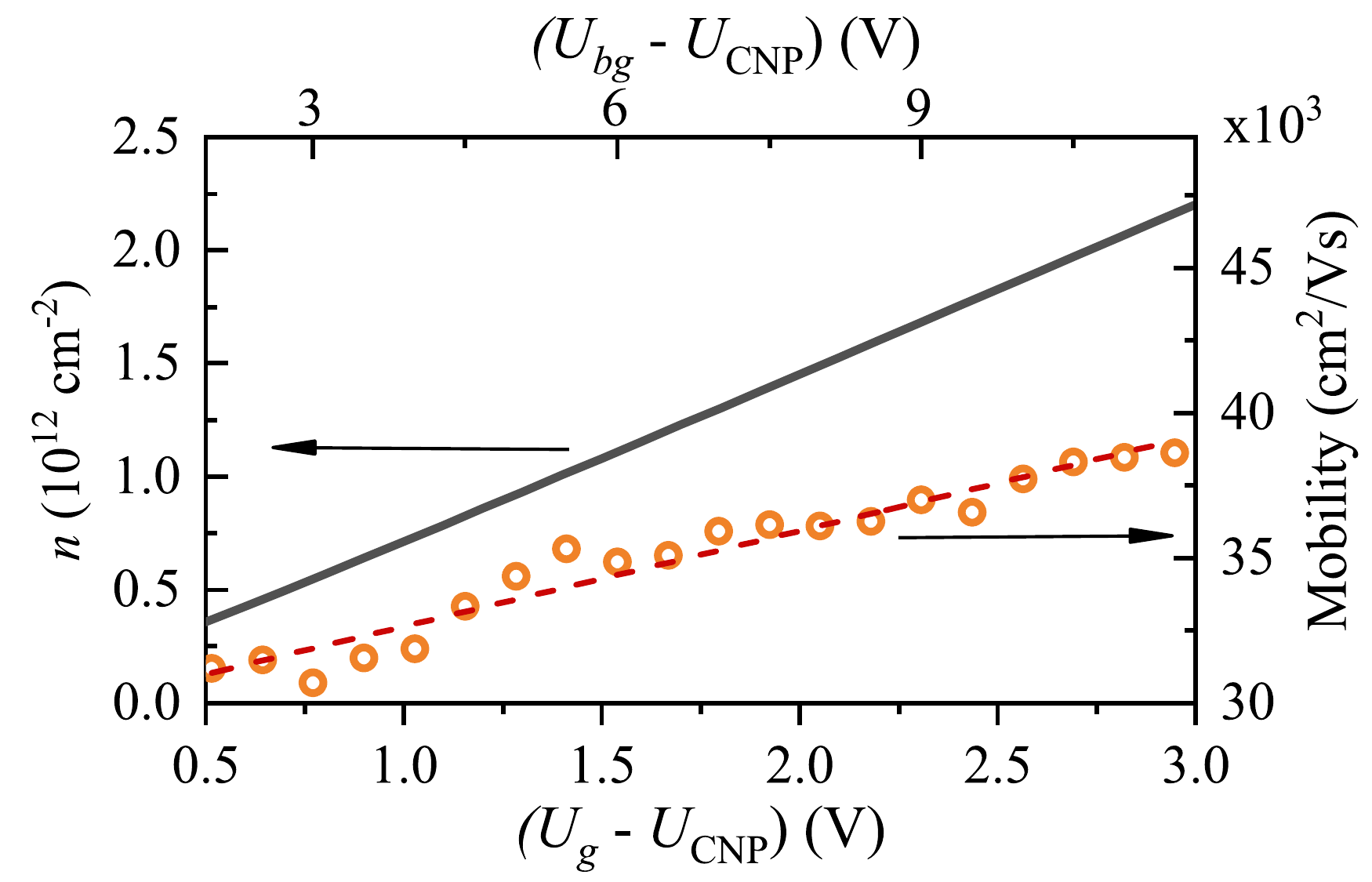}\par
\caption{ Electrical characterization of the samples. Ambipolar resistance versus back gate (a), and top gate for \textit{C}1 cavity (b) of A-DGG3.1 structure. We also show resistance versus inverse swing voltage $(1/(U_{g}-U_{\rm CNP})$ that allows extraction of the resistance of the contacts (a) and the resistance of total ungated part of samples for \textit{C}1 cavity (b). \textit{C}1 cavity carrier density, \textit{n} and mobility as a function of the gate voltage (c): lower horizontal-axis corresponds to the top gate bias and upper horizontal-axis is for back gate bias.}\label{Fig2}
\end{figure}

In the experiments, we manipulated the electron  concentration in the channel by gate voltages to create active and passive plasmonic  regions with high and low electron concentration, respectively. Specifically, we always biased the gates in such a way that in given experiment configuration, only one type of gates were positively biased (electrical n-type doping) and all other types of gates  were set at voltages granting the CNP condition of the graphene layer. 

In this way by successive biasing the different gates, we defined four types of grating gate structures (\textit{C}1, \textellipsis, \textit{C}4) with cavities of length 0.5, 0.75, 1.0, 1.1~\textmu m, each composed of six fingers with highly doped n-type cavities (up to $ \sim 2~\times 10^{12}$~cm$^{-2}$) separated by six regions at the CNP condition ($n_2 \sim 0.1~\times 10^{12}$~cm$^{-2}$ for both electrons and holes at room temperature). The parameters of these structures are given in Table~\ref{tab:table1}. 
The example of typical carrier density distribution in the experiments on cavity \textit{C}2 is shown in Fig.~\ref{Fig1}(c). 
We also show an example of the potential distribution in the case of application of 100~mV drain-to-source voltage (Fig.~\ref{Fig1}(c)).

The example of the results of electrical characterization (sample A-DGG 3.1) is shown in Fig.~\ref{Fig2}. The multiple sweeps of gate voltages have shown only minimal effects of hysteresis (see Fig.~\ref{Fig2}), and granted good reproducibility of all the experimental results. 

In Figs.~\ref{Fig2}(a) and~\ref{Fig2}(b) we also present the sample resistance as a function of the inverse swing voltage, $1/(U_{g} - U_{\rm CNP}$), allowing to determine the parasitic resistance, R$_{0}$, i.e., the sum of the contact resistances and resistances of ungated CNP biased parts of each sample.  One can see that as expected from the sample architecture, the biasing of the front gates change the resistance in a much smaller degree than the back gating. The resistance change is roughly proportional to the ratio of the length of the biased gate ($\times$6) and the total length of the channel, as given in Tab.~\ref{tab:table1}. Knowing the total parasitic resistance $R_{0}$ we can extract sample averaged conductivity $\sigma = (L_{ch}/W)/(R-R_{0})$, where $L_{ch}$ is the total channel length, $W$ is the average channel width, $R$ is the total resistance.
The carrier mobility $\mu_i$ can be calculated from equation  $ e \mu_i(n_i+p_i)  = \sigma_i$ (see  ~\cite{Satou2016, Ryzhii2007PlasmaHeterostructures} for details).
The electron and hole density $n_i$ and $p_i$ (index $i=1,2$ numerates regions with high and low conductivity, respectively) as a function of the back gate voltage, can be calculated using the parallel-plate capacitor model with the correction by the quantum capacitance of graphene. 
Figures~\ref{Fig2}(c) shows carrier density and mobility as a function of the front gate voltage (lower horizontal scale) and back gate voltage (upper horizontal scale). As can be seen, the mobility is over 30,000~cm$^{2}$/Vs in a whole investigated range indicating the excellent quality of h-BN-encapsulated graphene.
\begin{table*}
\caption{\label{tab:table1} Samples and different plasmonic cavities parameters.}
\begin{ruledtabular}
\begin{tabular}{lcccc}
\textrm{Cavity}&
\textit{C}1&
\textit{C}2&
\textit{C}3&
\textit{C}4\\
\colrule
Structure/sample & A-DGG3.1 & A-DGG3.2 & A-DGG3.1 & A-DGG3.2\\
Thickness of top h-BN layer (nm)	& 32 & 20 & 32 & 20\\
CNP (V)	& +0.15 & -0.12 & +0.10 & -0.06 \\
Biased cavity length (\textmu m) & 0.5 & 0.75 & 1.0 & 1.5\\
Total channel length (\textmu m) & 26.5 & 24.0 & 26.5 & 24.0\\
 d$_{1}$ and d$_{2}$ (\textmu m) & 0.5 / 2.0 & 0.5 / 1.0 & 0.5 / 2.0 & 0.5 / 1.0\\
 Channel width (average) (\textmu m) & 4.9 & 1.325 & 4.9 & 1.325\\
 d$_{3}$ and d$_{4}$ (\textmu m) & 2.0 / 0.5 & 1.0 / 0.5 & 2.0 / 0.5 & 1.0 / 0.5\\
\end{tabular}
\end{ruledtabular}
\end{table*}

\subsection{\label{sec:TDS}Time-domain spectroscopy experiments}
 We use in our experiment grating-gate-based structures. Such structures have been used  to study  the plasmon excitation in two-dimensional electron gas by incident THz wave  since  the  pioneering  works of Allen and Theis~\cite{Allen1977a, Theis1977a}. In particular,  recently the grating gate 
structures were   used in graphene plasmons experimental studies~\cite{Olbrich2015PRB}.  Also, the electromagnetic simulation of plasmon excitation by THz wave in graphene with an asymmetric grating gate (which is very similar to the structure measured in our paper) was recently presented in~\cite{Fateev2019PRA}. These works show that  the plasmon resonances can be excited in grating gate graphene  structures   in reflectivity configuration and  can be used to obtain the THz radiation absorption.

\begin{figure}[!ht]
         \centering
             (a)\includegraphics[width=0.94\linewidth, bb = 0 0 450 225]{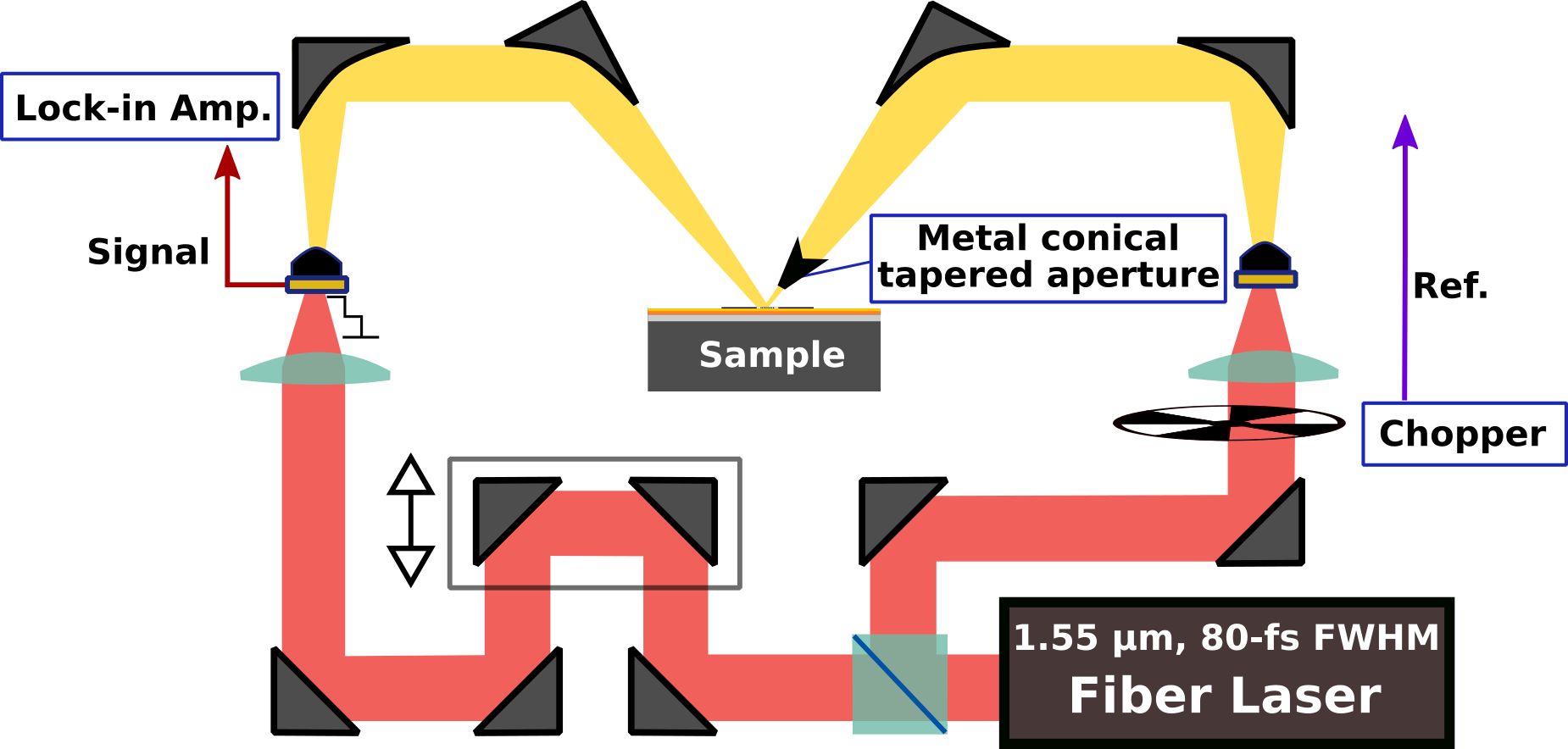}\par 
             (b)~~\includegraphics[width=0.9\linewidth, bb = 0 0 500 312]{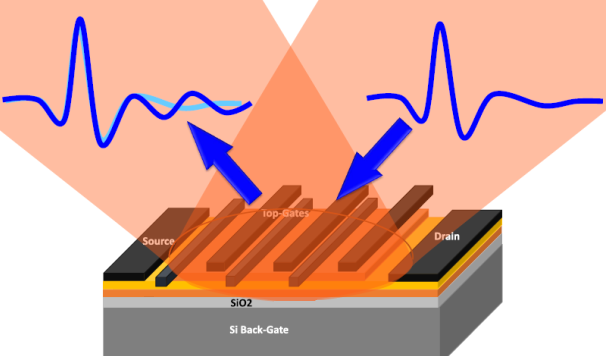}\par
             (c)~~\includegraphics[width=0.9\linewidth]{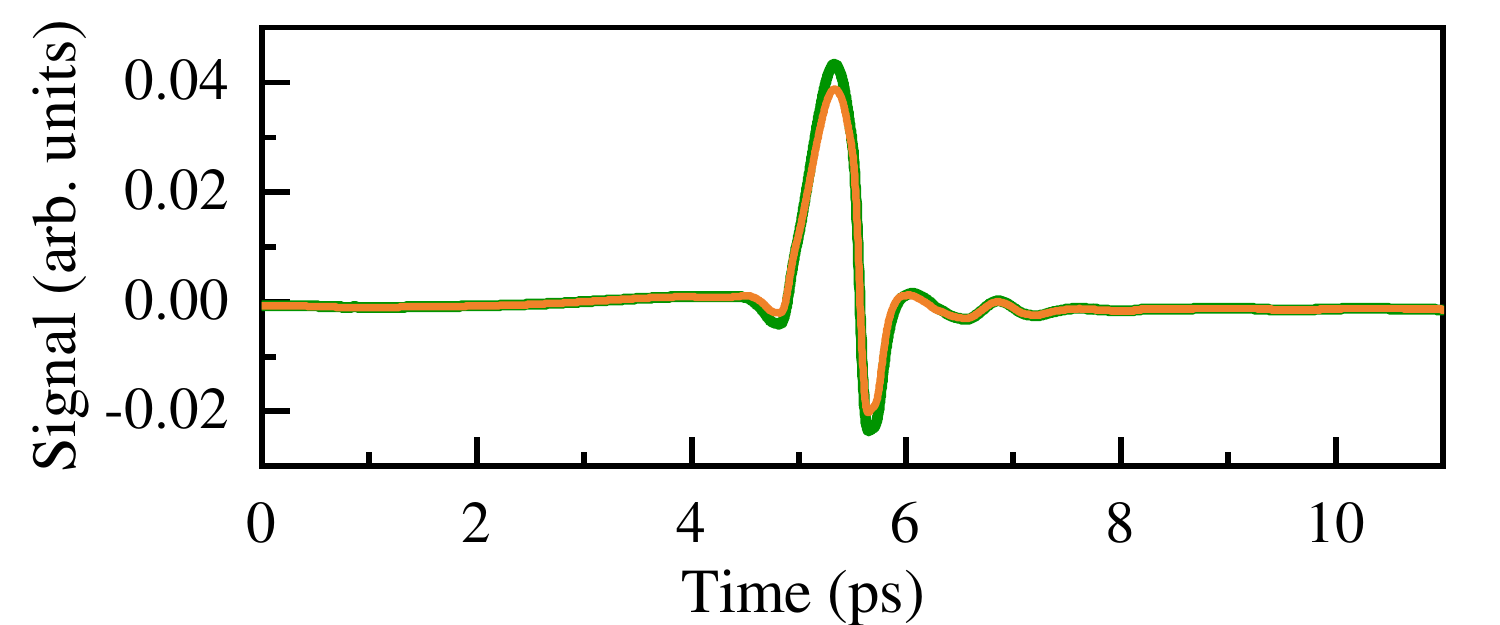}\par 
             (d)~~\includegraphics[width=0.9\linewidth]{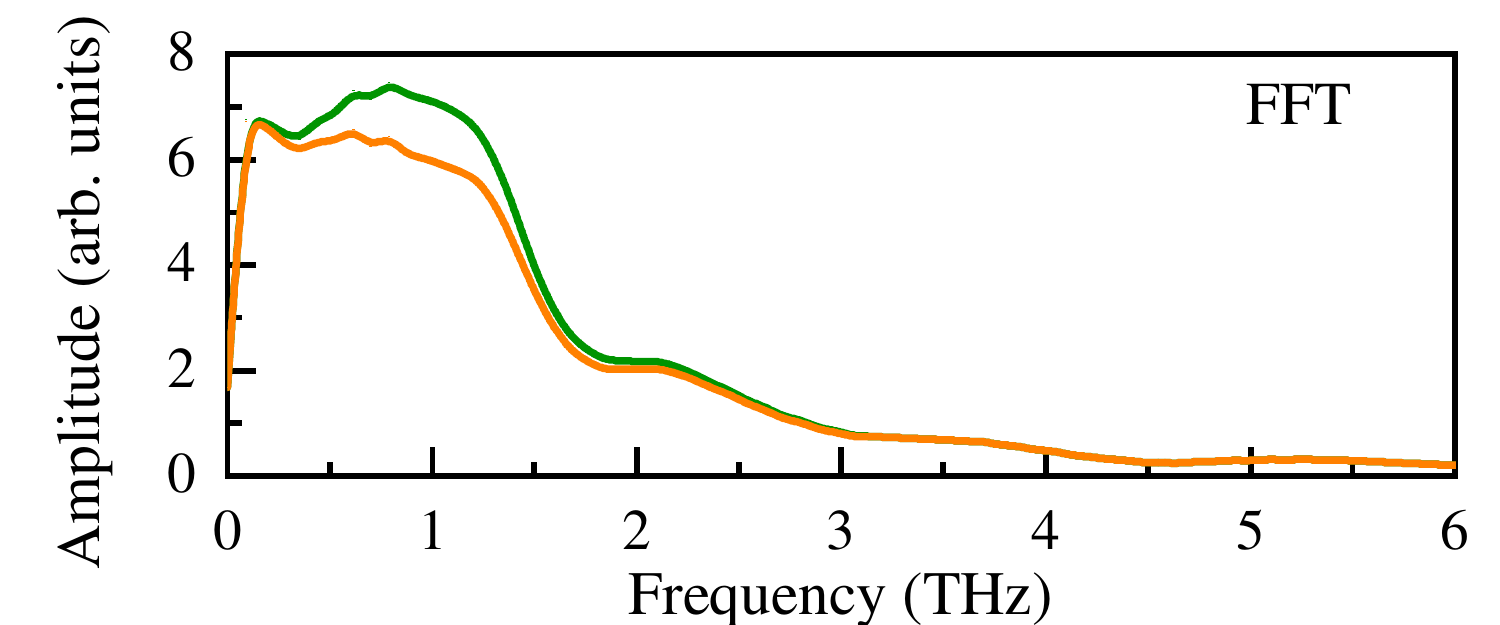}\par 
     \hfill{}
     \caption{Measurement setup. (a) Terahertz time-domain spectroscopy (THz-TDS). (b) Total reflection configuration. The incident THz radiation pulse was linearly polarized. The sample stage was rotated to obtain the radiation polarization perpendicular or parallel to the drain-source direction. (c) Typical time trace for samples with (orange) and without gate biasing (green) in the measurement system are shown in upper panel. Spectrum are obtained by FFT transformation of the same signal in lower panel}
\label{Fig3}
\end{figure}

Experimental set-up configuration is shown in Fig.~\ref{Fig3}(a). 
THz time-domain spectroscopy (THz-TDS) was employed to measure the changes in the temporal profiles of the THz pulses transmitted through the graphene plasmonic cavities. 
For the generation of broadband THz waves, a mode locked erbium-doped femtosecond fiber laser with a pulse repetition rate of 80~MHz was used. The pulse width was approximately 80~fs (full width at half maximum) and the central wavelength was 1550~nm. 

The pulsed laser beam was focused onto a biased, low-temperature-grown InGaAs/InAlAs THz photoconductive antenna (TERA15-TX, Menlosystems). The antenna had a gap spacing of 100~\textmu m and a bias of 15~V was applied with an average optical power of 20~mW. 
The linearly polarized THz wave emitted by the antenna was collected and refocused onto the sample from 45$\degree$ oblique angle (and subsequently onto the THz detector) using off-axis parabolic mirrors. A small 250~\textmu m-diameter hole at the end of a tapered aperture of a conical shape was placed close to the sample. We have checked that this aperture was not destroying the polarisation of the incoming beam. THz radiation was detected using a second low-temperature-grown InGaAs/InAlAs THz photoconductive antenna made of the 25~\textmu m dipole and 10~\textmu m gap excited with an average optical power of 17~mW (at 1550~nm). 

All the TDS measurement were performed in a time window of 12~ps with a temporal resolution of 6.67~fs. Examples of typical time traces and spectra obtained with and without electrical doping are shown in Figs.~\ref{Fig3}(c), the orange and green lines respectively. The time traces in the time window of 12~ps were used in the calculation of the Fourier transform spectra. Typical Fourier spectra (obtained from Fig.~\ref{Fig3}(c)) in a linear scale are shown in Fig.~\ref{Fig3}(d), confirming an effective frequency bandwidth going from 0.1~THz going up to above 4~THz with a rather high signal-to-noise ratio (above $10^4$) in the whole range. In the rest of the work, we restrict the presentation for time and frequency ranges – where the real differences between the measurement with and without electrical doping (with and without strong positive gate polarization) are well seen.
The sample stage was rotated to align the THz radiation polarization perpendicular or parallel to the grating fingers direction.

In the first part of experimental studies, we measured the spectra for zero drain-to-source voltage ($U_{d}$ = 0). As mentioned above, the spectra were always recorded with only one gate voltage biased away from CNP and with all other gates biased at the CNP. In the second part of experimental studies,  $U_{d}$-dependent measurements were performed.

We used the reflection configuration, as shown in Fig.~\ref{Fig3}(b). The THz pulse was transmitted twice through the h-BN, graphene and thin h-BN/SiO$_{2}$ dielectric layers. Due to highly doped low resistivity conditions of the Si substrate with the 45$\degree$ oblique angle incidence, the interface gave total reflection. When the THz pulse traveled through such the path, it reflected and transmitted through at the interfaces between graphene and the h-BN buffer layer and between the h-BN buffer layer and the SiO$_{2}$ layer. 

Such a multi-layered vertical structure makes a superposition of the reflected pulses at each interface in its temporal pulse profile. Since the thicknesses of the h-BN buffer layer ($\sim$ 40~nm) and SiO$_{2}$ layer ($\sim$ 90~nm) are far smaller than the wavelength of the THz radiation, the artefacts and distortions caused by the multiple reflections on the temporal pulse profile are negligibly small (the round-trip delay time from/to the top graphene surface will be $\sim$ 3~fs which is almost two-orders shorter than the THz pulse width). Therefore, what is measured by the TDS is the total pulse after its double passage throughout the aforementioned multi-layered sample. 
\begin{figure}[!ht]
         \centering
             \includegraphics[width=1\linewidth]{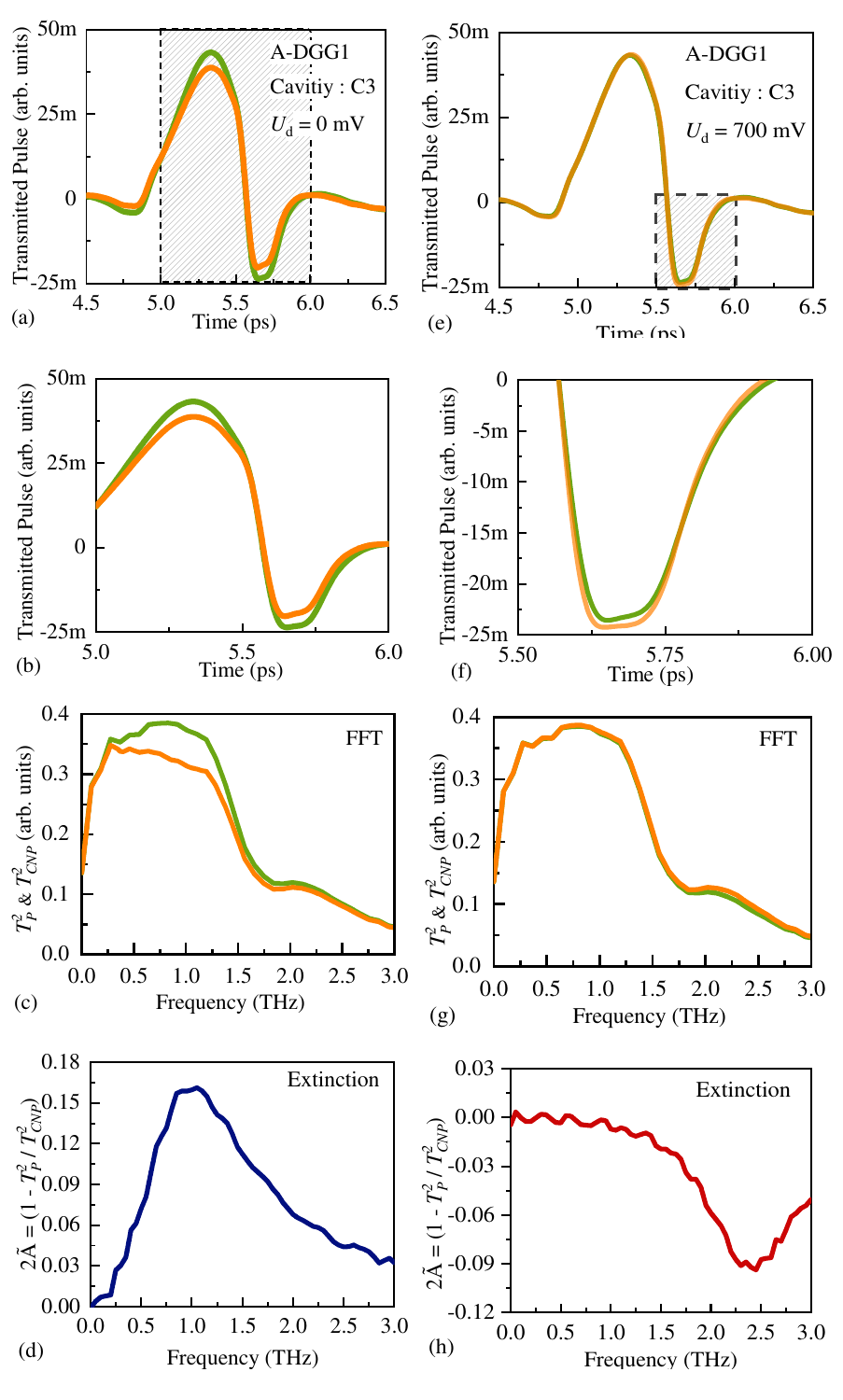}\par 
     \caption{Typical time traces obtained for A-DGG1 sample with (orange) and without gate biasing (green). Left panels (a)-(c) experiments without drain-source biasing ($U_{d}$~= 0), right panels (e)-(g) the results obtained with a strong drain biasing at $U_{d}$~= 700~mV.
     Panels (b) and (f) are magnification of the shadowed regions in the panels (a) and (e) respectively. One can observe that the drain-source biasing inverses the order of magnitude of orange and green traces in the time traces and spectra row data. 
     Panels (d) and (h) show spectra obtained 
     by Fourier transform for time traces with and without gate biasing, respectively.}
\label{Fig4}
\end{figure}



Some typical representative data are shown in Fig.~\ref{Fig4}. In each measurement we registered two time traces and calculated spectra: i) the first $T_{\rm CNP}^2$, with all gates biased the way ensuring CNP condition of the whole graphene channel and ii) the second $T_P^2$ with only one type of the grating gate \textquoteleft strongly biased\textquoteright (up to $U_{g} - U_{\rm CNP} \sim$ 3~V) while keeping all other parts of the sample at the CNP condition. Then we calculated $2 \tilde A~=(1-T_{P}^2/T_{\rm CNP}^2)$. As it will be shown later $2 \tilde A$ contains mainly information about the extinction. The left panels in Fig.~\ref{Fig4} are examples of some data (time traces, Fourier transform and resulting extinction) obtained without any drain-source biasing ($U_{d}$~= 0), whereas the right panels are the results obtained with a rather strong drain-source biasing at $U_{d}$~= 700~mV. Even if the differences in traces are difficult to see at the first glance– they are very well resolved in the experiment because of, aforementioned, very high signal to noise ratio (above $10^{4}$ in whole presented spectral range). 
The shadowed regions in Fig.~\ref{Fig4} were magnified in the neighbor panels to show the regions where the differences can be observed.
First, one can clearly see that the time traces (and spectra) without and with strong biasing (electrical doping) show some well-resolved experimental differences. Second, one can observe that the application of additional drain-source biasing (700~mV) inverses the order of the blue \textquotedblleft without carriers\textquotedblright and red “with carriers” traces – suggesting \textquotedblleft negative transmission\textquotedblright. This very important point will be systematically investigated and discussed in the next parts of this paper.

Generally, the experimental spectra $T_{P}^2$ and $T_{\rm CNP}^2$ contain information about THz transmission, reflection and absorption. The key parameter that governs the  underlying physics in an array of conducting strips is the ratio of the conductivity  in the active region to the light velocity: $2\pi \sigma_1/ c \sqrt{\epsilon}$ (see Refs.~\cite{Mikhailov1998,Ju2011}  and discussion in  Sec.~A of the Supplementary material).  Here $\epsilon$ is the dielectric constant of the surrounding media. 

For the case, when this parameter is small,
\begin{equation}
\frac{2\pi \sigma_1}{ c ~\sqrt{\epsilon} } \ll 1,
\label{parameter}
\end{equation}
the transmission coefficient is  approximately given by 
\be
T\approx 1-A,
\ee
where $A \ll 1 $ is the absorption coefficient (see  Supplementary Material). The quantity, which is measured in the experiment, is proportional to $T^2,$ because incoming beam goes through the array of the conducting strips, reflects from the metallic substrate (mirror), and it goes again through the array of the conducting strips. We get
\begin{equation}
T^2\approx (1-A)^2 \approx 1-2 A.
\label{TT}
\end{equation}

As mentioned above the quantity that we interpret in the experiment is $(1 - T_{P}^2 /T_{\rm CNP}^2 )$, where $T_{\rm CNP}^2$ is the spectra registered with whole sample at CNP condition and  $T_{P}^2$  is the spectra obtained with only one of the grating gates ``biased''. Using Eq.~\ref{TT} one can show that measured quantity is proportional to $2A$. In fact, correction to the transmission coefficient in our experiment is smaller because the size of the beam $S_b$ is bigger than the samples grating area $S_{g}$. The measured absorption is given by $2 \tilde A  = 2 A  S_b/S_{g} . $   
 
The coefficient $A$ can be simply    expressed  in terms of radiation-induced  correction  to the dissipation in the  system, $\delta P$:
\be
A = \frac{ \delta P}{ {\cal S}}= \frac{8\pi\delta P}{ c \sqrt{\epsilon} E_0^2}.
\label{A_1}
\ee 
Here $E_0$ is the amplitude of the incoming radiation and ${\cal S} = c \sqrt{\epsilon} E_0^2/8\pi$ 
is the  time-averaged   radiation Pointing vector for linearly-polarized wave.   
Using  Eq.~\eqref{TT} and Eq.~\eqref{A_1}, we obtain   
$(1 - T_{P}^2 /T_{\rm CNP}^2)  \approx  2 \tilde A \propto \delta P.$
This is the quantity $2 \tilde A $
which we plot in all experimental figures. 
Importantly, this quantity is  proportional to the  dissipation in the channel---the property, which we will use for the theoretical interpretation of our results. 

We emphasize that in addition to dissipation in the channel,
there also exist radiation losses. However, a detailed analysis performed  in Ref.~\cite{Mikhailov1998}  for a similar structure showed that the radiation losses are small when condition \eqref{parameter} is satisfied. In particular, the rate of radiation attenuation 
($\Gamma$ in the notation of reference \cite{Mikhailov1998}) is small compared to the  momentum relaxation rate $\gamma$
provided that  the  condition \eqref{parameter} is fulfilled. 
Therefore, it is enough for us to analyze the rate of dissipation in the channel and this radically simplifies the calculations.

\section{\label{sec:results} EXPERIMENTAL RESULTS}
\subsection{\label{sec:zero_condition}Resonant plasmons at zero drain current conditions}

\begin{figure}[!ht]
\includegraphics[width=1\linewidth]{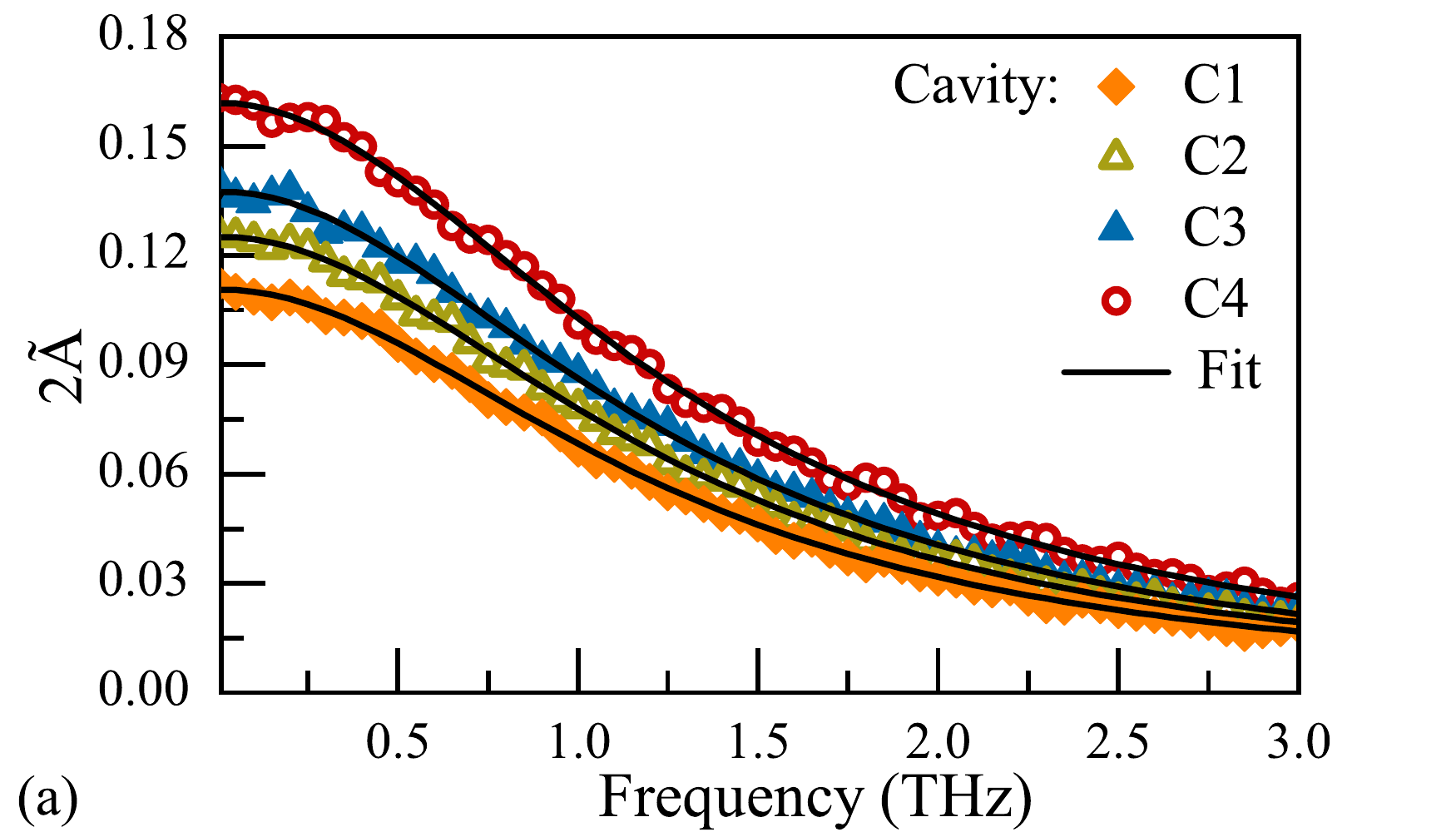}\par 
\includegraphics[width=1\linewidth]{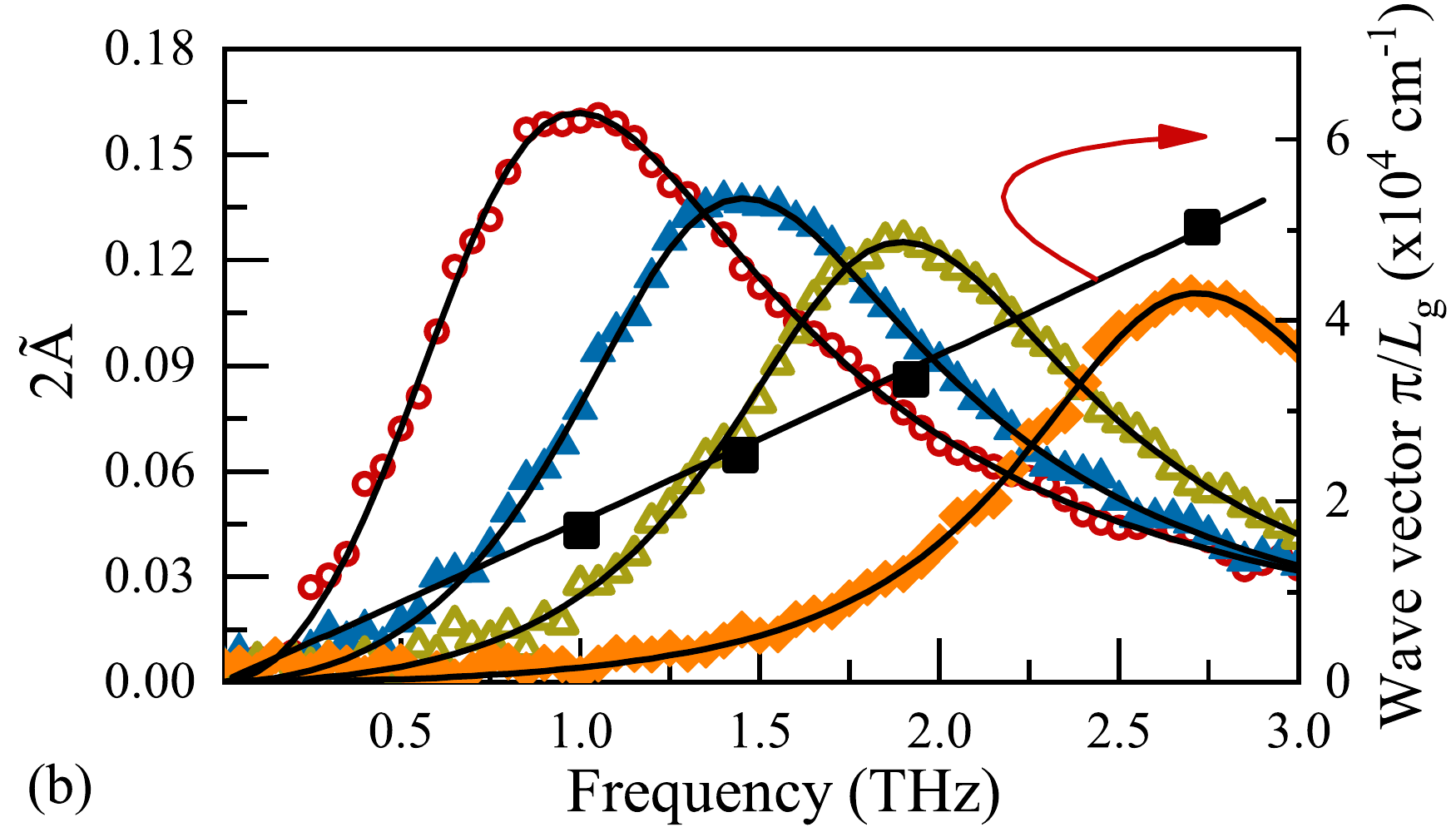}\par
\caption{Gate length dependent extinction spectra of the graphene structures for $U_{d}$ = 0~V with incident light polarized parallel (a) and perpendicular (b) to the gates fingers. The measured data points and lines are the fits by the Drude model fit for parallel polarization and the damped oscillator model for perpendicular polarization. 
Black solid squares are a maximum of experimental resonant plasma frequencies as a function of the wave vector $q = \pi/L_1$ . 
Black straight line through these dots is the best linear fit (right hand scale).}
\label{Fig5}
\end{figure}

In Fig.~\ref{Fig5} we show the results of some systematic measurements with zero drain voltage/current applied to the structures ($U_{d}$ = 0~V).

Comparison of the  measurements results for two different polarizations of the THz radiation: parallel and perpendicular to the gate fingers are shown in Figs.~\ref{Fig5}(a) and \ref{Fig5}(b) respectively. This comparison is done for the same electrical doping of cavities ($U_{g} - U_{\rm CNP} \sim$ 3~V). One can see that, in the parallel polarization case, the extinction spectra are characterized by the Drude-like response with a monotonic decrease of the absorption with frequency~\cite{Ju2011}. In contrast, in the perpendicular polarization, a completely different line shape was observed with a pronounced resonant absorption peak shifting to higher energies for narrower gate fingers. 

In the case of the polarisation parallel to the stripes  the   dissipation is due to the  Drude absorption and measured extinction spectra can be fitted by using the Drude formula:

\begin{equation}
\label{eq:1}
\tilde A \propto \frac{\gamma_1 }{\gamma_1^2 +\omega^2},
\end{equation}
where $\gamma_1=1/\tau_1$  is the plasmon relaxation rate in active area and  coefficient in this equation is frequency-independent.
The fitted values of the amplitude and plasma frequency varied for each cavity, while the variation of $\tau _{1}$ was fairly small: 0.12 $\sim$ 0.13~ps.

Likewise, using the results of the supplementary materials, we fitted the frequency dependencies of the extinction spectra for the perpendicular polarization case by formula of the so-called damped oscillator model~\cite{Ju2011, Yan2012a, Yan2012b, Yan2013}

\begin{equation}
\label{eq:2}
\tilde A \propto  \frac{\gamma_1}{\gamma_1^2 + (\omega-\omega_1^2/\omega)^2 } \approx   \frac{\gamma_1/4}{\delta\omega^2 +\gamma_1^2/4}. 
\end{equation}
where $\omega _{1}$ is the resonance frequency in the active region with high conductivity and $\delta \omega =\omega-\omega_1.$
The lines in Fig.~\ref{Fig5} correspond to theoretical calculations. The calculated curves agree quantitatively very well with the  extinction spectra registered in the experiments. 

It is worth noting that the radiation with parallel polarization does not lead to redistribution of charge even in the case of 
non-zero \textit{dc} current, so that plasmonic effects do not show up for such a polarization.  

In Fig.~\ref{Fig5}(b) we show also the frequency as a function of the wavevector ($q = \pi/L_{1}$) where $L_{1}$ is the single finger gate width of the active region with high conductivity (right hand scale). One can see that the resonant frequency scales almost linearly with $q$ determined by the length of the grating finger, suggesting that in the investigated structures the grating fingers act as independent cavities with 2D gated plasmons – with dispersion relation close to $\omega = s_1 q$, with $s_1 \sim 3 \times 10^6$ m/s. 
The important thing is that plasma wave velocities are in all regions higher then Fermi velocity in graphene. As a consequence in our experimental situation, the carrier drift velocities in all regions are smaller than the plasma wave velocities.

Summarizing, the results from Fig.~\ref{Fig5} show  that: (i) the metallic gates are close enough to 2DEG to grant dispersion very close to linear - strongly gated plasmon dispersion, and (ii)  it is not the grating period, but the single finger dimension that determines the plasmons wave vectors.  

\begin{figure}[!ht]
\includegraphics[width=0.9\linewidth]{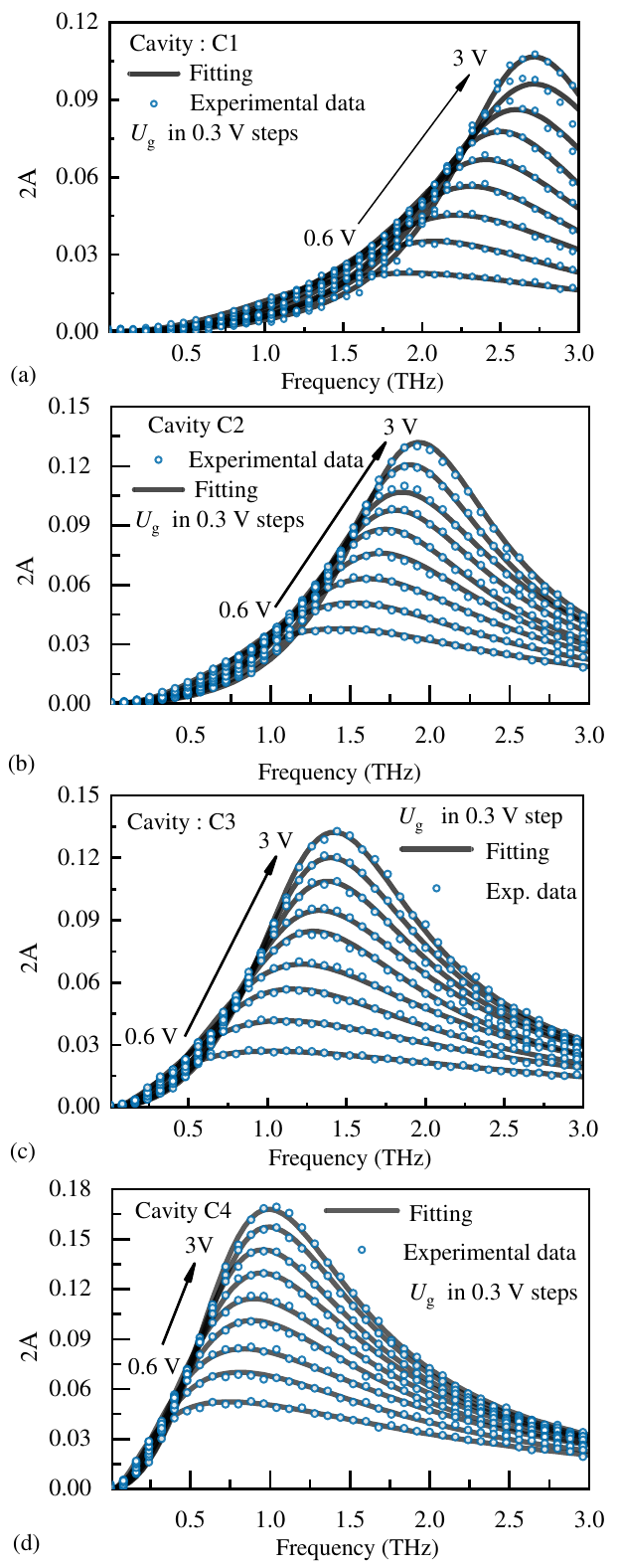} 
\caption{ Gate-voltage-dependent extinction spectra of the graphene structures for $U_{d}$ = 0~V with incident light polarized perpendicular to the gate fingers. Panels (a), (b), (c), and (d) presents results for \textit{C}1, \textit{C}2, \textit{C}3, and \textit{C}4 cavities, correspondingly. The experimental data are marked as dots, and continuous lines are the results of fits using standard physical models of gated 2D plasmons in graphene.}
\label{Fig6}
\end{figure}

In Fig.~\ref{Fig6} we show the results of the study of the plasmon resonances, as a function of the gate voltage (2D carrier density).  As in all experiments, the gate voltage-dependent extinction spectra are shown while tuning the only one cavity gate at the time  and keeping all other parts of the sample at the charge neutrality condition. The gate voltage-dependent data show a clear blue-shift and increase of the strength of the absorption peak frequency with increasing carrier density (increasing gate voltage). As shown in Figs.~\ref{Fig6}(a)-~\ref{Fig6}(e), the fitting curves obtained using 
the damped oscillator model  agree very well with the measured extinction spectra. 

\begin{figure}[!ht]
\includegraphics[width=1\linewidth]{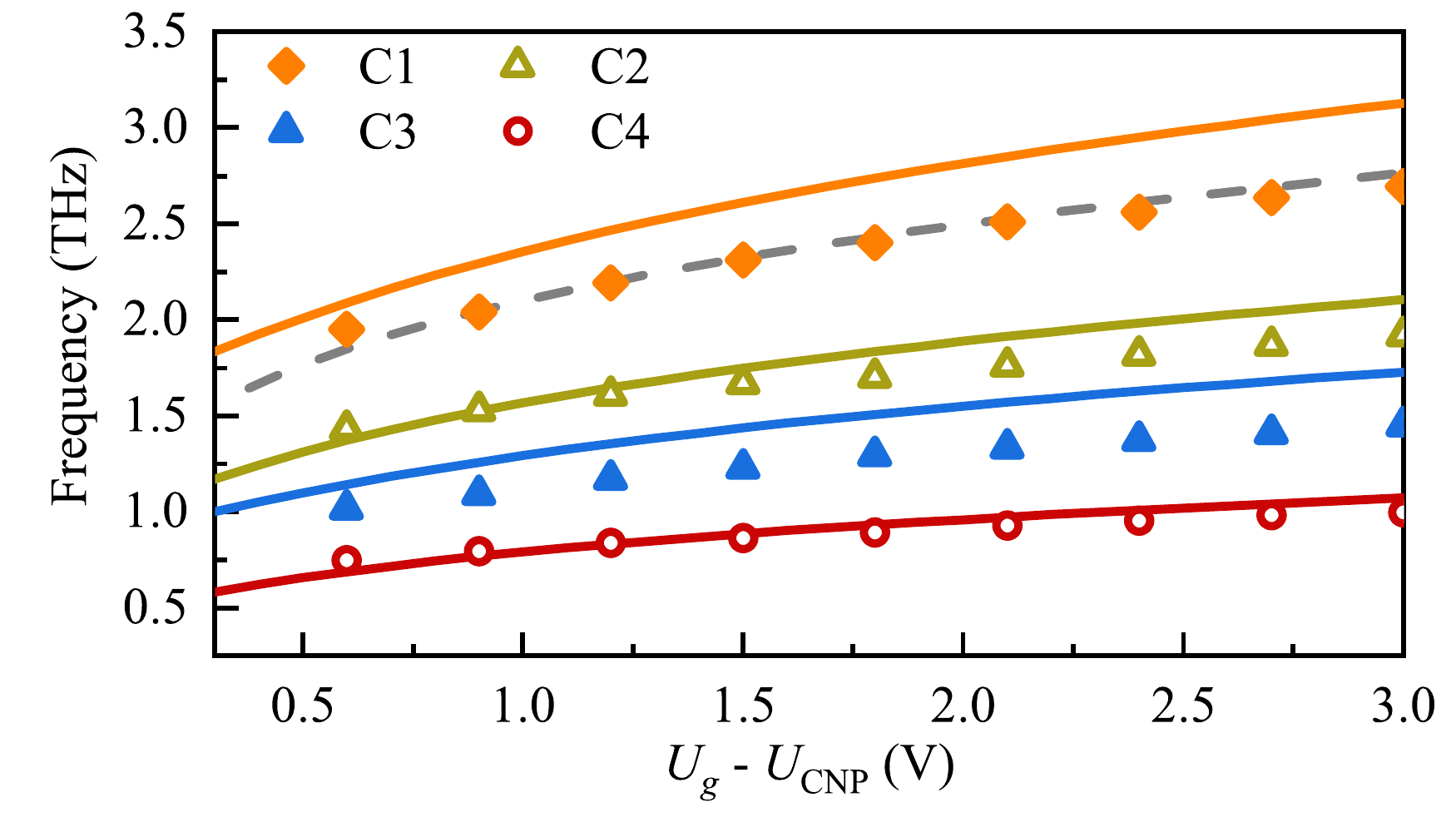}
\caption{Gate voltage dependence of the extracted resonance frequency.
Continuous lines are results of calculations according Eq.~\eqref{eq:omega} with corrections to fringing and plasma leakage phenomena- see supplementary materials for more details. The dotted line is $\nicefrac{1}{4}$~power dependence as a function of the gate bias, typical for plasmons in graphene.}
\label{Fig7}
\end{figure}

Figures~\ref{Fig7} shows the gate-voltage dependences of the extracted plasmon frequencies. One can see that a sample with a shorter cavity length has a higher frequency at a given gate voltage.
Figure~\ref{Fig7} also shows theoretical plasmon frequencies, as functions of the gate voltage calculated as
\begin{equation}
\label{eq:omega}
\omega_{q} = \sqrt{\frac{4 e^2 T\ln[2+2\cosh(\varepsilon_F/T)]}{\hbar^2 \varepsilon q[1+\coth(qd)]}}q,
\end{equation}
where $q$ is the plasma wave vector, $d$ is the thickness of the top h-BN layer given in Tab.~\ref{tab:table1}, $\varepsilon$ = 4.5 is the dielectric constant of  h-BN layers~\cite{Laturia2018DielectricBulk}, $\varepsilon_F$ is the  Fermi energy, $T$~=~300~K is the temperature.
At sufficiently high electron concentration ($\varepsilon_F \gg T,$), Eq.~(\ref{eq:omega}) simplifies to the well known form~\cite{Ryzhii2006TerahertzHeterostructures,Ryzhii2007PlasmaHeterostructures}:

\begin{equation}
\label{eq:3}
\omega_{q}= \frac{s q}{\sqrt{qd[1+\coth(qd))]}}.
\end{equation}
Here
\be
s=\sqrt{ \frac{4 e^2 \varepsilon_F  d}{\varepsilon \hbar^2} },
\label{ss}
\ee
is the plasma wave velocity ($s=s_1$ in the active region, and $s=s_2$ in the passive region). 
The factor $q d  [1+\coth(qd)]$ describes deviation from the linear dispersion and is important when the thickness of the top h-BN layer is comparable with the gate lengths ($L_{1}$ or $L_2$).

In the calculations,  we have taken into account the quantum capacitance correction~\cite{Satou2016}, fringing and plasma leakage as predicted by phenomenological model described later in the work and in the supplementary materials. One can see that the theoretical calculations made without any fitting parameters reproduce relatively well the absolute values and the functional gate voltage dependencies of the plasma frequency for all four cavities, close to $\nicefrac{1}{4}$~power dependence. The small discrepancies may result from the uncertainty of h-BN layer thickness determination (precision $\sim 10\%$) and unknown exact value of the dielectric function. In fact in the literature one can find $\epsilon$ in the range from 3.5 to 5 \cite{PhysRev.146.543, Dean2010BoronElectronics, Yu2013InteractionCapacitance, Laturia2018DielectricBulk}. Also in our calculations graphene Fermi velocity was assumed carrier density independent, where in principle it may change in the range, between 1.3~$\times 10^{6}$ to 1.0~$\times 10^{6}$~m/s in our experimental conditions \cite{Hwang2012FermiModification, Yu2013InteractionCapacitance}.  

It should be stressed however, that independently of small numerical discrepancies, the observed plasma frequency follows $\nicefrac{1}{4}$~power dependence on gate voltage (carrier density) typical for plasmons in graphene - as shown by dotted line in Fig.~\ref{Fig7} (see also Fig.~\ref{Fig6}(b)
in supplementary materials).  

THz plasmon resonances in graphene have been already observed in micro-ribbons~\cite{Ju2011, Tomasino2013a, Strait2013a}, rings, disks~\cite{Yan2012a, Yan2012b, Yan2012c,Daniels2017NarrowGraphene} and grating-gate structures~\cite{Bylinkin2019Tight-BindingGraphene}. However, it is worth to mention that:  i) here we report the first experimental observations of  such plasma standing waves with 100\%  gate tunable frequency  and ii) the plasmon related extinction coefficient is going up 19\%.

To summarize this part, the scaling behavior of plasmon frequency versus the plasmon wave vector and the gate voltage shown in Figs.~\ref{Fig5}, \ref{Fig6} and \ref{Fig7} were successfully calculated using standard physical models of gated 2D plasmons in graphene~\cite{Ryzhii2006TerahertzHeterostructures, Hwang2007DielectricGraphene, Ryzhii2007PlasmaHeterostructures}. 
 
They clearly confirm the existence of 2D plasmons oscillating at THz frequencies in our graphene/h-BN nanostructures and allow an unambiguous attribution of the observed resonances to 2D plasmons in the individual fingers of the grating-gate defined cavities \textit{C}1,..., \textit{C}4. 

\subsection{\label{sec:dc_case}Plasma resonances in the presence of the \textit{dc} current}
The most important part of the present work concerns the experimental investigations of the influence of the \textit{dc} drain-source bias $U_{d}$ on the identified earlier THz 2D plasmons resonances.

In Fig.~\ref{Fig4} one can see examples of experimental time traces and their Fourier transform for cavity \textit{C}2 (1~\textmu m) with and without applied drain bias. The results: first ($T_{\rm CNP}^2$) with all the gates \textquoteleft unbiased \textquoteright (set at the CNP) and the second with only \textit{C}2 cavity gate \textquoteleft biased\textquoteright~at $\sim$~3~V \textendash~$(T_{P}^2)$ are plotted. Comparing the results with and without drain-source biases, one can clearly see that in the range of plasmonic resonances the order of the traces with and without gate biasing is inverted after switching on the drain-source voltage (700~mV). First for zero drain bias, when the cavity \textit{C}2 is filled with carriers \textendash the outgoing light intensity is lower than the incoming one (plasma resonant absorption). For high drain voltage (700~mV), in contrast, the registered signal is higher when the cavity is filled with carriers (amplification or emission).This is the reason why in the last right panels data plotted as $2 \tilde A$ show negative values.  

\begin{figure}[!ht]
\centering
\includegraphics[width=0.8\linewidth]{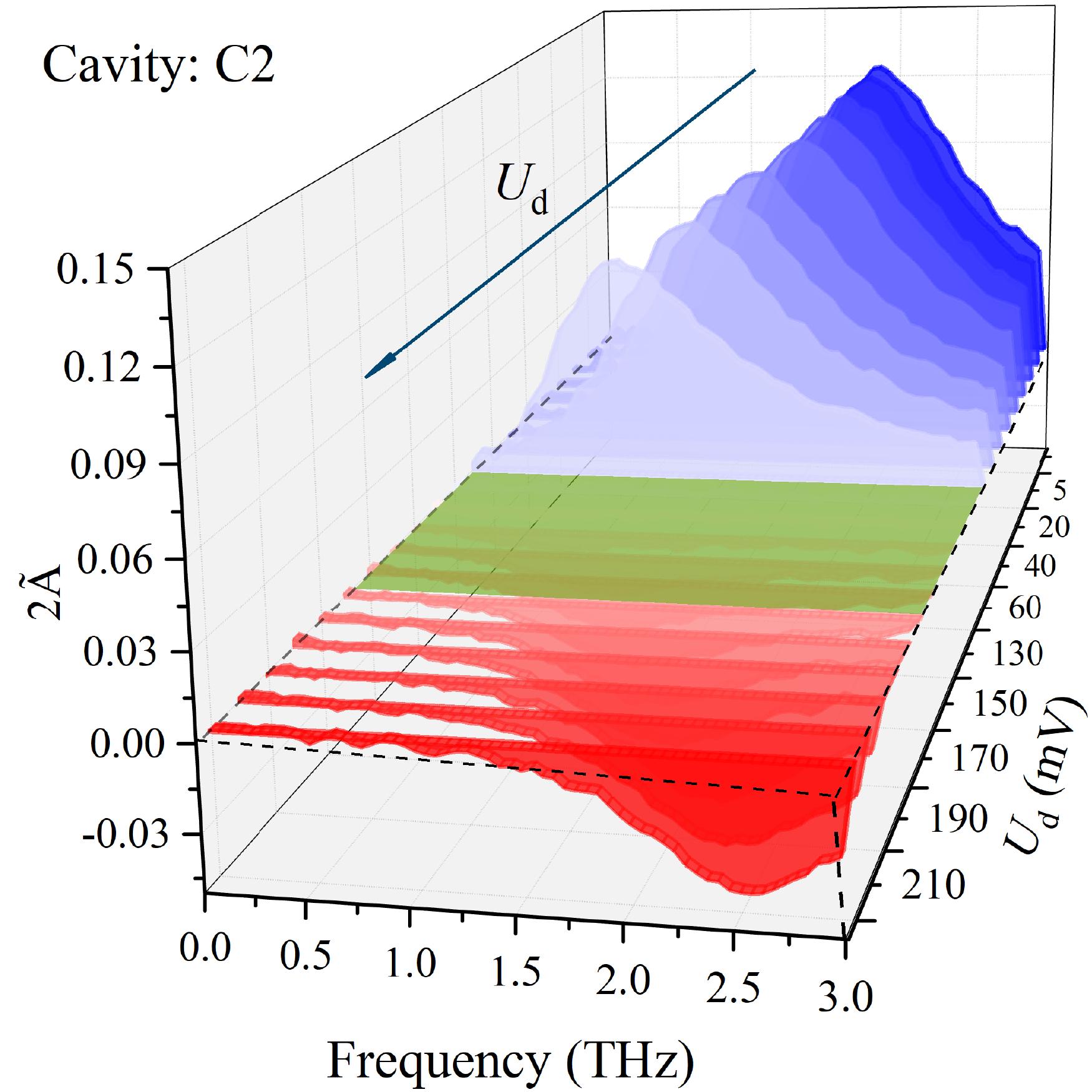}\par 
\caption{Drain bias dependent spectra for the \textit{C}2 plasma cavity. Blue color marks absorption; green  zone is the gap between absorption and emission and the red colour corresponds to amplification. Arrow indicates raising drain bias.}
\label{Fig3D}
\end{figure}

To further  investigate  this intriguing behavior, we performed very systematic measurements versus drain bias for all four cavities. 
In Fig.~\ref{Fig3D} we show 3D plot illustrating the typical drain bias dependence of the extinction (\textit{C}2 plasma cavity). With increase of the \textit{dc} current, a strong red shift of the resonant THz plasmon absorption, a region of complete transparency to incoming radiation, followed by the amplification and blue shift of the resonant plasmon frequency are observed. 

\begin{figure}[!ht]
\centering
\includegraphics[width=1\linewidth]{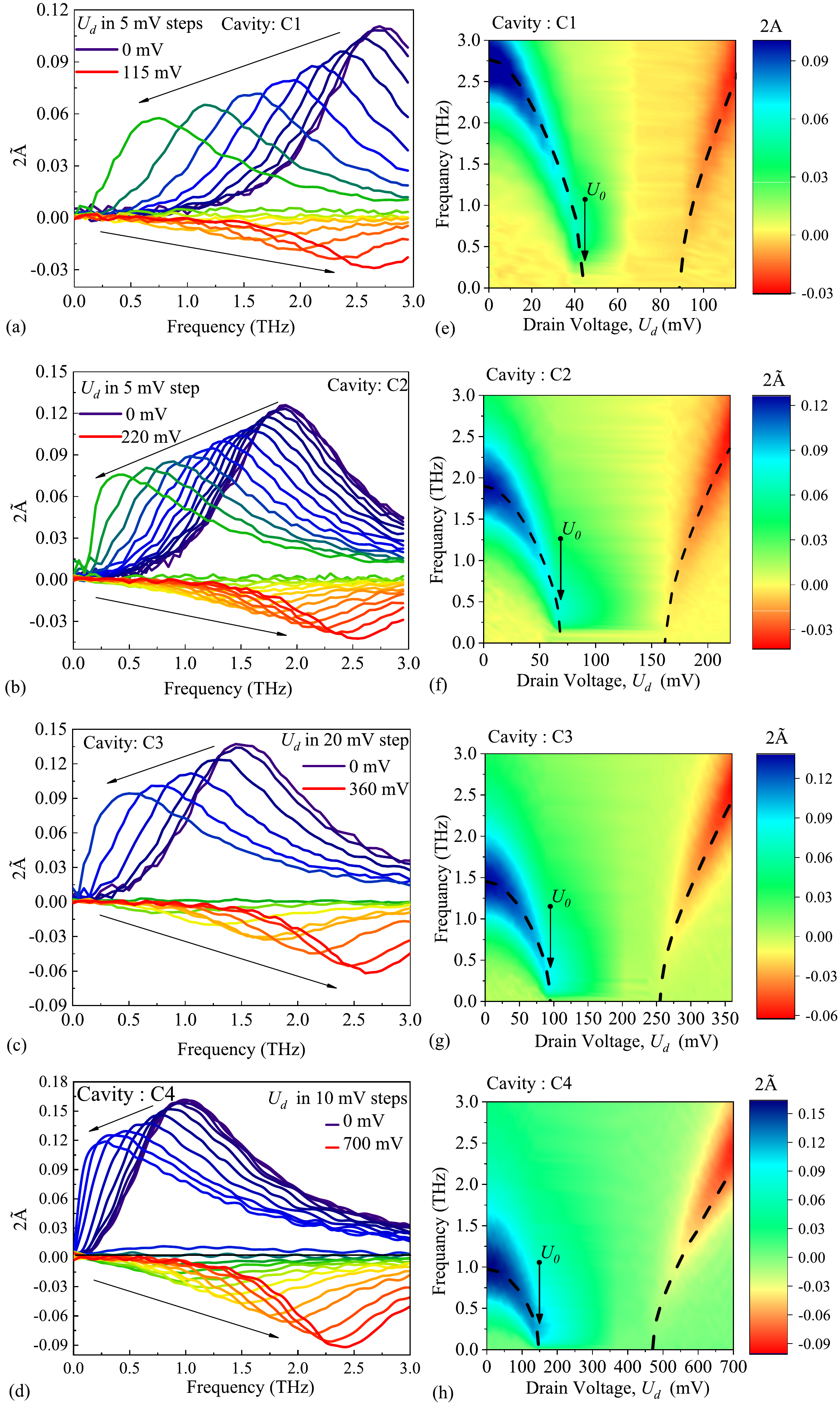}\par 
\caption{Drain-bias-dependent properties of the \textit{C}1, \textit{C}2, \textit{C}3, and \textit{C}4 plasma cavities. The left-hand panels show the spectra (a), (b), (c), and (d), correspondingly. The right-hand panels show the contour plots (e), (f), (g), (h), correspondingly. The dotted lines on contour plots are results of the fit with a phenomenological formula Eq.~\ref{fit_omega}~–as described in Sec. III.}
\label{Fig8}
\end{figure}

In Fig.~\ref{Fig8} we collect all results in the form of the 2D plots - keeping the colour code as in the Fig.~\ref{Fig3D}. 
As $U_{d}$ increases,  the absorption peak clearly shifts to lower frequencies along with a noticeable reduction of plasmon resonance amplitude. Then the absorption completely vanishes, as seen for example in the measured extinction spectra for cavity \textit{C}1, being zero at $U_{d} \sim$ 45~mV up to $U_{d} \sim$ 90~mV [see Fig.~\ref{Fig8}(a)]. In this drain voltage range, the plasmonic device \textit{C}1 becomes perfectly transparent to the incoming THz radiation within the entire experimental bandwidth. It is important to stress that, here we report the first experimental observation of such a strong Doppler shift of plasma resonances in graphene and the transparency behavior over a relatively wide frequency range (0.1 to 3~THz). 
With increasing $U_{d}$ beyond this transparency regime, a resonant ``negative absorption''  feature appears in the extinction spectra and shifts with increasing drain bias towards higher frequencies (noticeable blue shift). 

Throughout all of our experiments, the data from all cavities \textit{C}1, \textellipsis, \textit{C}4 were quite similar with the amplification threshold voltage increasing with the cavity length $L_{g}$ and important widening of the transparency range (from about 50~mV up to  300~mV). 

\begin{figure}[!ht]
\centering
\includegraphics[width=1\linewidth]{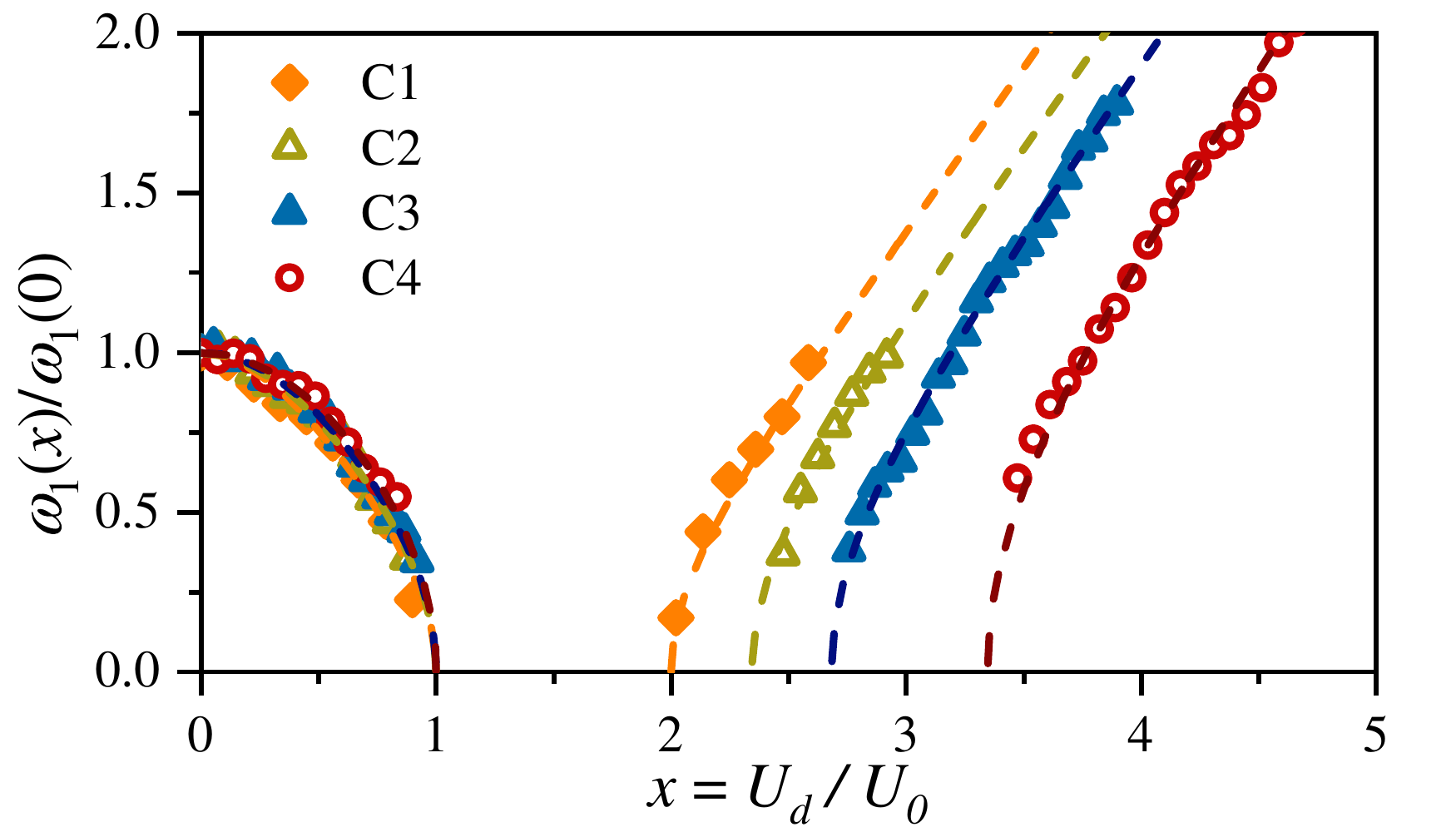}
\caption{The experimental results showing resonant frequency versus drain voltage bias in normalized scales. Dotted lines show results of  calculations according to   Eq.(\ref{fit_omega}).}
\label{FigXXX}
\end{figure}

The results presented in Fig.~\ref{Fig8} show some universal behaviour. This is visualised when we present them in re-scaled mode. In Fig.~\ref{FigXXX}, the vertical axis is the frequency divided by its value at $U_d = 0$ and the drain voltage (the horizontal axis) is normalized by the drain voltage $U_0$ at which the plasmon frequency tends to zero (see Fig.~\ref{Fig8}). Very good quantitative agreement both at low and at high currents can be obtained by fitting experimental curves  as
\be
\frac{\omega_1^2(x)}{\omega_1^2(0)} =1+\zeta + x^2 - \sqrt{\zeta^2 + 4x^2 (1+\zeta) }
\label{fit_omega}
\ee
where $\omega_1(x)$ stands for position of the plasmonic resonance, $x={U_d}/{U_0}$,  $\omega_1(0)$ is the experimentally measured value of the plasmonic frequency at zero current. $U_d$ is the voltage across the structure, $U_0$ is the voltage corresponding to onset of the transparency window, and $\zeta$ is a fitting parameter. 
The values of $\zeta$ corresponding to solid lines in Fig.~\ref{FigXXX} are presented in Table~\ref{tab:table2}.
The unified universal behavior of  frequency at low currents (for $x<1$) can be obtained by taking in Eq.~\eqref{fit_omega} the limit $\zeta \to \infty,$ which yields     
\be
\frac{\omega_1^2(x)}{\omega_1^2(0)}=1-x^2=1- \left(\frac{U_d}{U_0}\right)^2, 
\label{fit0omega}
\ee
The similar functional dependence of the plasmon frequency versus electron drift velocity of 2D electron gas in GaAs/AlGaAs grating gate structures (Eq.~\ref{fit_omega}) was obtained in Ref.~\cite{Mikhailov1998}. However, the physical model used in this work can not be applied to interpret our results. Validity of the physical model of Ref.~\cite{Mikhailov1998} for our structures and analogy used to derive Eq.~\ref{fit_omega} will be discussed  in more details in the following section and in the Supplementary materials.

\begin{table*}
\caption{\label{tab:table2} Properties of different plasma cavities.}
\begin{ruledtabular}
\begin{tabular}{lcccc}
\textrm{Cavity Name}&
\textit{C}1&
\textit{C}2&
\textit{C}3&
\textit{C}3\\
\colrule
Structure/sample & A-DGG2 & A-DGG1 & A-DGG2 & A-DGG1 \\
\colrule
$L_1$ (\textmu m) & 0.5 & 0.75 & 1.0 & 1.5  \\
$L_2$ (\textmu m) & 3.5 & 3.0 & 3.0 & 2.25 \\
$U_0$ (mV) & 45 & 69 & 92 & 143  \\
$U_T$ (mV) & 90 & 159 & 267 & 458  \\
$s_1$ ($10^8$ cm/s) & 3.2 & 3.2 & 3.1 & 3.1  \\
$s_2$ ($10^8$ cm/s) & 1.6 & 1.6 & 1.6 & 1.6  \\
$f_1 = \omega_1/2\pi$ (THz) & 2.76 & 1.90 & 1.45 & 0.97  \\
$f_2 = \omega_2/2\pi$ (THz) & 0.11 & 0.23 & 0.24 & 0.32  \\
$\zeta$  & 1.5 & 2.3 & 3.1 & 5.1  \\
$n_1$ ($10^{12}$ cm$^{-2}$) & 2.0 & 2.0 & 2.0 & 2.0  \\
$n_2$ ($10^{12}$ cm$^{-2}$) & 0.1 & 0.1 & 0.1 & 0.1  \\
\end{tabular}
\end{ruledtabular}
\end{table*}

\section{Phenomenological description of the obtained results }
\subsection{Plasmonic crystal model.}

\begin{figure}[!ht]
\centering
\includegraphics[width=0.45\linewidth]{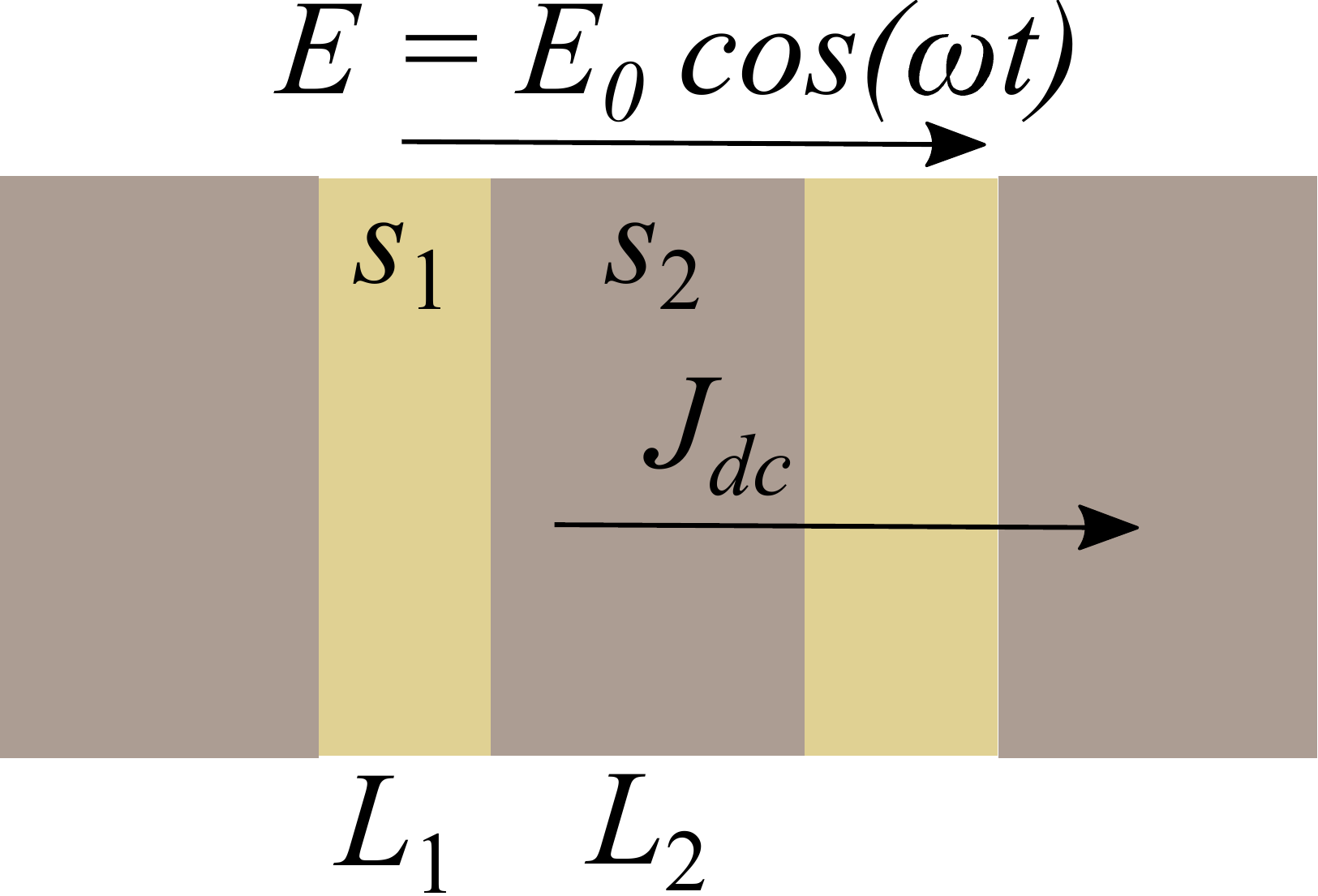}\par 
\caption{1D plasmonic crystal driven  by \textit{dc} current  in the oscillating field  of THz radiation.   }\label{fig1new}
\end{figure}

In phenomenological interpretation, we consider a plasmonic crystal composed of two different regions with lengths $L_1$ and $L_2$, plasma velocities $s_1$, $s_2$, carrier density $n_1$, $n_2$ and drift velocities ${\cal V}_1$, ${\cal V}_2$ (see Fig.~\ref{fig1new}). We assume that $n_1 \gg n_2$ and refer region ``1'' as active and region ``2'' as passive.  The plasma waves velocities $s_1$ in the electrically doped active cavities (for gate voltage 3~V) were calculated from experimental data shown in Fig.~\ref{Fig7}, using Eq.~\ref{eq:3}. The results are shown in Table ~\ref{tab:table2}. One can see that for all cavities $s_1 \sim 3.2 \times 10^8$ cm/s. Because the plasma velocity is proportional to the ($1/4$) power of the carrier density(see Eq.~\ref{ss}), the plasma velocities in CNP passive cavities can be estimated as $s_2 \sim s_1/2 \sim 1.6 \times 10^8$ cm/s. 
The drift velocities in the electrically doped regions (active regions) ${\cal V}_1$ are by more than one order of magnitude lower than these in the CNP regions. The ratio (${\cal V}_2/{\cal V}_1 \sim  n_1/n_2 \sim 10$) of the drift velocities stems from the  \textit{dc} current continuity condition ($n_1 {\cal V}_1 = n_2 {\cal V}_2$). We also take into account in this subsection  that damping rates in the passive and active regions can be different. We assume, however, that they have the same order of magnitude: $\gamma_1 \sim \gamma_2.$

Importantly, plasma wave velocities are in all regions higher than Fermi velocity in graphene. As a consequence, in our experimental situation,  the carrier drift  velocities in all regions are smaller than the plasma wave velocities.

Therefore the most important theoretical challenge is to find the physical process/mechanism in which the amplification can take place for drift velocities below the plasma wave velocity. In this work, we will develop a  phenomenological theory which shows that indeed, amplification is possible for ${\cal V}_1<s_1.$ 
This theory shows a possible way of physical interpretation and gives  universal dependence of the plasmonic frequency on the current in a sense that the only controlling parameter is ${\cal V}_1/s_1$.

\begin{figure}[!ht] 
\centering
\includegraphics[width=0.6\linewidth]{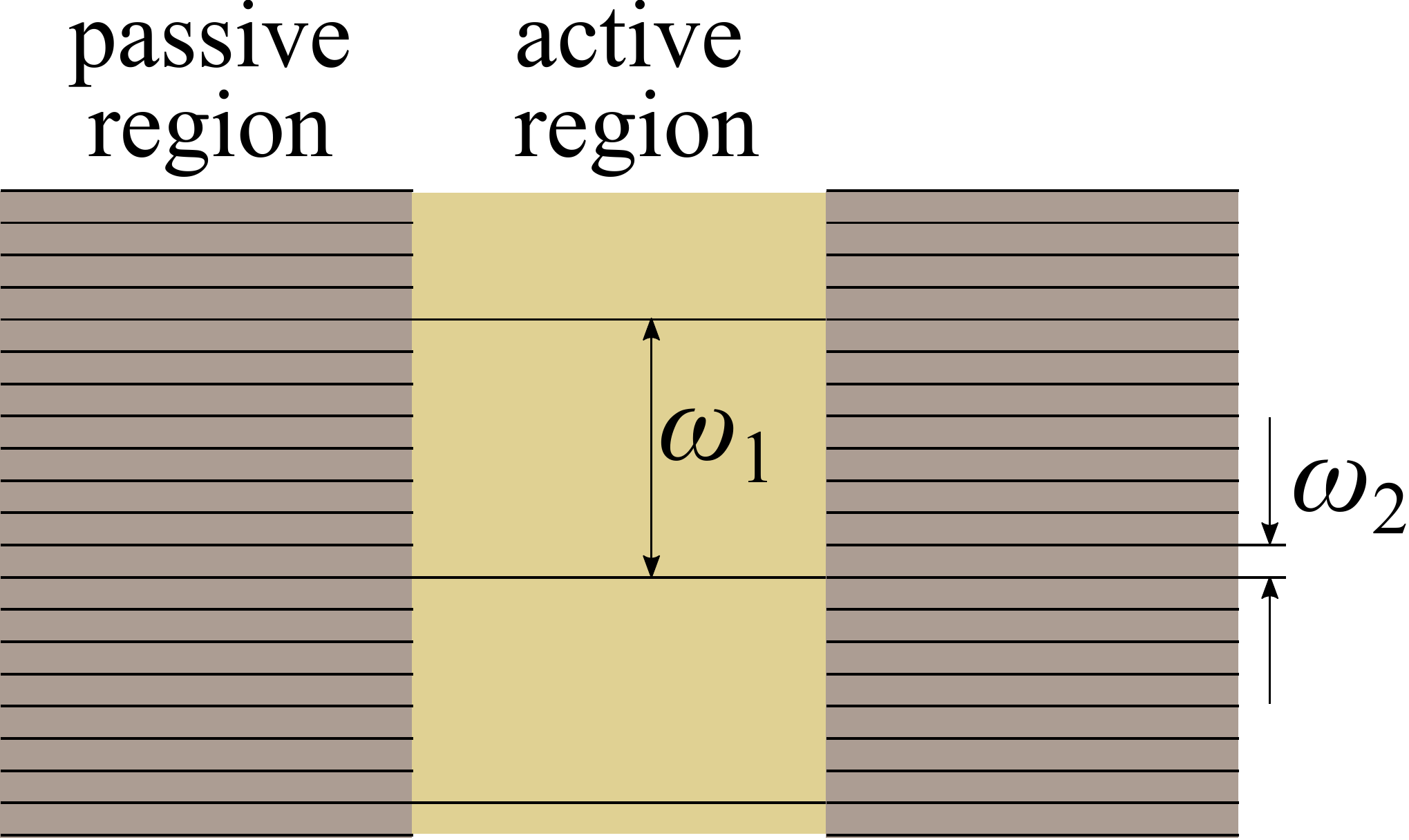}\par 
\caption{Active regions of the plasmonic crystal with large fundamental frequency are separated by passive regions with small frequency}\label{fig0new}
\end{figure}

Let us consider the structure shown in Fig.~\ref{fig1new}.  In the absence of dissipation, such structure represents an example of 1D plasmonic crystal, with the plasma wave   spectrum, $\omega(k),$  which describes   allowed and stop bands  and  should be found from the following  dispersion equation   \cite{Kachorovskii2012a}:
\be
\cos(kL)\!=\!\cos(\omega T_1)\cos(\omega T_2)\!-\!Z\sin(\omega T_1)\sin(\omega T_2).
\label{bands}
\ee
Here $k$ is the plasmon quasimomentum,  $T_{1}=L_1/s_1,$  $T_{2}=L_2/s_2$ are the plasmon transit times, and $Z = (s_1^2+ s_2^2)/(2 s_1 s_2)  $ is the mismatch parameter. For simplicity, we consider theoretically  only the case $s_1 \gg s_2,$ when $Z \gg 1.$
In this case, allowed bands  have negligible widths and can be found from equation $\sin(\omega T_1)\sin(\omega T_2)=0,$ which describes   plasmonic oscillations  in the active regions with the fundamental frequency  $\omega_1=\pi s_1/L_1 $  and in the passive regions with the fundamental  frequency $\omega_2=\pi s_2/L_2 $ (see Fig.~\ref{fig0new}).

Let us assume that  this  structure is illuminated by  radiation with the large wavelength,  $\lambda \gg L_1,~\lambda \gg L_2,  $  so that  the electric field  of radiation is approximately  homogeneous. 
This field excites plasmonic oscillations   both in active and passive regions. The plasmonic resonances occur, 
when  frequency of the external field equal to  $n\omega_1$ and $m \omega_2,$ where $n$ and $m$ are  the integer numbers.  
Equation \eqref{bands} can be easily generalized for  realistic dissipative   case $\gamma_1\neq 0, \gamma_2 \neq 0,$ when spectrum 
of the plasmonic crystal  acquires also an imaginary part: $\omega(k)=\omega'(k) +i \omega''(k).$ Corresponding dispersion equation is quite       cumbersome and we do not present it here.   However, the physical  results obtained from this equation can be understood based on simple physical consideration.   

We  assume  that  
\be s_1/L_1 \gg \gamma_1 \sim \gamma_2 \gg s_2/L_2.  \label{condition}\ee
In this case, resonances in the passive region strongly broaden and  overlap. Physically, this means that plasma waves rapidly decay along the passive region, which, in turn, implies that different active regions are disconnected  at plasmonic frequencies. This is in an excellent agreement with experimentally observed independence of different active regions. On the other hand, different active regions are  connected  at zero \textit{dc} frequency in a sense that \textit{dc} current flows through  the  system.     

Let us clarify this point in more detail.  The oscillations  decay into passive region as $\propto \exp(-\gamma \delta x/s_2)$ (here $\delta x$ is coordinate counted from the border between regions).  Under the condition \eqref{condition},   this exponent  at  $\delta x=L_2$ becomes $\propto \exp(-\gamma L_2/s_2) \ll 1.$  Hence,  the  coupling of oscillations in the neighboring active regions is exponentially  small   and one can reduce the problem to  calculation of excitation in  a single active strip with proper boundary conditions. A detailed discussion of such an approach is presented in the Supplementary material. For zero current, the proper conditions look
\be \delta j_1 (0)=\delta j_1(L_1)=0,  \label{BC}\ee
which means that the current is fixed on both sides of the active region.  Such conditions allow us for detailed quantitative  description of plasmonic resonances in transmission coefficient. 
 
To check how inequalities Eq.~\eqref{condition} are fulfilled in our experimental situation we have to estimate  resonant frequencies in the passive and active plasma cavities. 
The length of CNP cavities do not vary much and for rough estimations, one can  take $L_2  \approx 3$~\micro m  which leads to plasma angular frequency as $\omega_2 =2  \pi f_2  \approx 1.6~$THz. One can see that the condition $\omega_1 \gg\omega_2$ (or, equivalently, $s_1/L_1 \gg s_2/L_2$) is relatively well fulfilled for all four experimental configurations (active cavities \textit{C}1,...\textit{C}4). 

\subsection{ Resonance frequencies and  broadening  of resonances for zero current.}
\begin{figure}[!ht]
\includegraphics[width=1\linewidth]{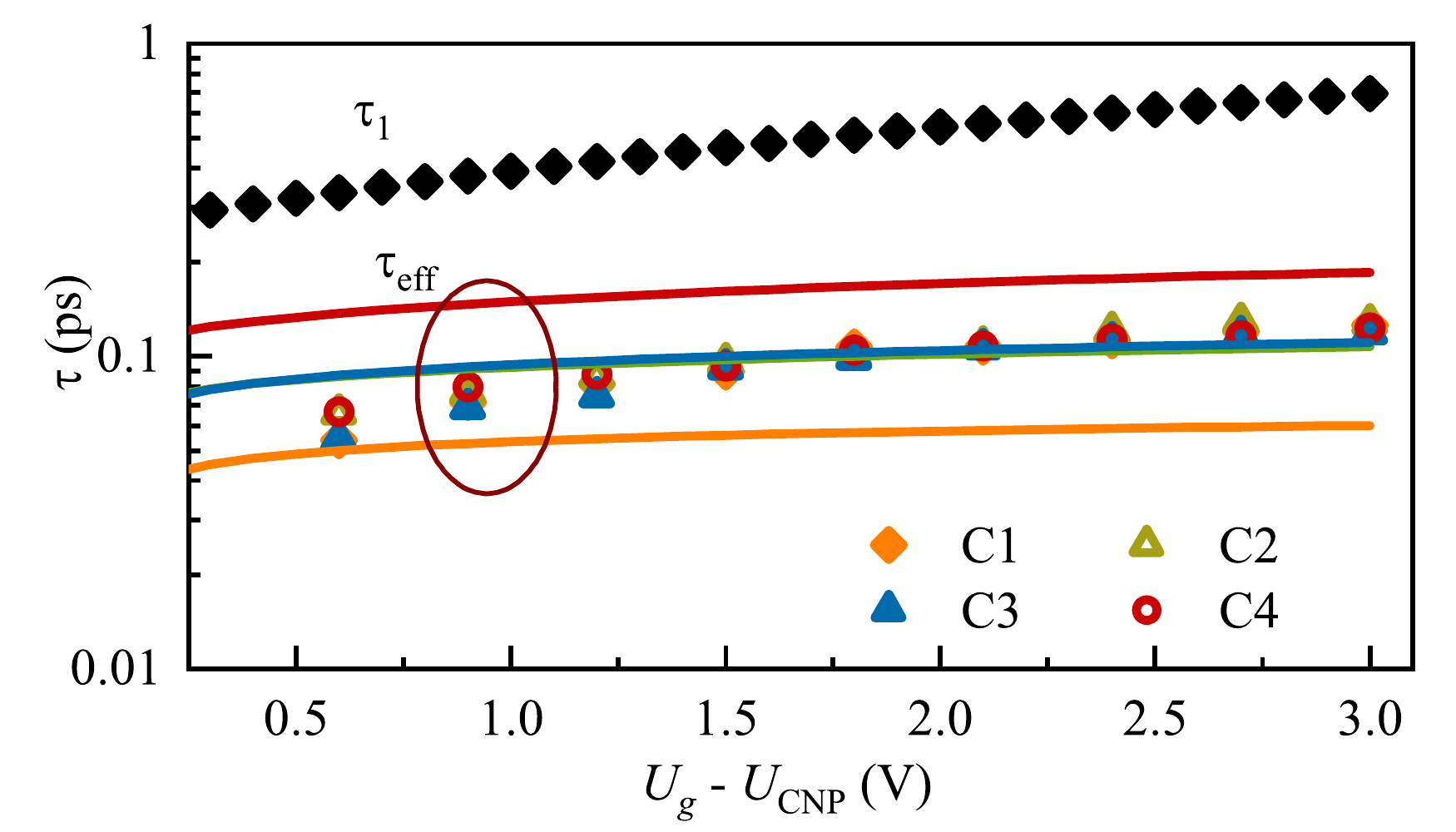}
\caption{Relaxation time as a function of the gate bias. $\tau_{1}$ is the momentum relaxation time in the active region  shown as the black diamonds, $\tau_{eff}$ is the plasmon relaxation rate calculated using Eq.~\eqref{leakage_main} (solid lines), color dots are experimental $\tau_{eff}$.}
\label{Fig_tau}
\end{figure}

Let us first consider the case of zero current. In this case, existence of the two plasma regions has two consequences - the plasma frequency is shifted toward lower frequencies and broadening appears (similar broadening  was predicted for a single gate problem 
in Ref.~\cite{Popov2018}).
These two phenomena are described by following equations  (see supplementary materials):
\begin{align}
& \gamma_{\rm eff}= \gamma_{1}+\frac{2s_1}{L_1} \ln\left( \frac{s_1+s_2}{s_1-s_2}\right),
\label{leakage_main}
\end{align}
where $\gamma_{1} = 1/\tau_1$, $\tau_1$ is the momentum relaxation time in the active region (in this subsection, we assume that $\gamma_1$ can differ from $\gamma_2,$ but have the same order of magnitude, $\gamma_1 \sim \gamma_2,$ so that  inequality \eqref{condition}  holds for both $\gamma_1$ and $\gamma_2$) 
\begin{align}
& \delta \omega_1 = \frac{\pi (\gamma_2-\gamma_1) s_1 s_2}{(s_1^2-s_2^2)\left[\pi^2+ \ln^2(\frac{s_1+s_2}{s_1-s_2}). \right]} ,
\label{real-part_main}
\end{align}
The effective width of the plasma cavities $L_1$ should also include  correction related to the fringing effect, that leads to the following correction of the plasmon frequency: 
\begin{align}
\delta \omega_{1}^{\rm fr} = \frac{2d}{L_{1}}\omega_{1}.
\label{d_omega_2}
\end{align}  
In Fig.~\ref{Fig_tau}, we show the momentum relaxation time $\tau_{1}$ determined from transport measurements together with plasmon relaxation time $\tau_{eff} =  1/\gamma_{eff}$, calculated using Eq.~(\ref{leakage_main}). We also show  experimental plasmon relaxation time determined from the fit of the resonances (see Fig.~\ref{Fig6}). One can see that leakage-induced contribution to the damping reduces the relaxation time by approximately one order of magnitude, and calculated relaxation times are in relatively good agreement with experimental data. 

This explains why the widths of plasma resonances are much larger than those  predicted from the transport experiments. 
It should be stressed that in the plasmonic crystal approximation developed here, the correction to the resonant frequency (Eq.(\ref{real-part_main})) is relatively small. Therefore the experiments in the absence of the drain bias can be interpreted taking  the standard  graphene plasmon description of independent  plasmon cavities Eq.~(\ref{eq:omega}), (\ref{eq:3}), (\ref{ss}). For a more detailed discussion of plasmons resonances width and corrections to plasma frequency see supplementary materials.

\subsection{Energy dissipation and amplification.} 
Next, we discuss the most important experimental  observation --- the  optical signal amplification  observed when the  drift velocity in the active region is smaller than  corresponding plasma wave velocity, $s_1$. 
To describe this phenomenon, we calculate dissipation in the channel under the drain biased condition and demonstrate that with increasing the current, it  decreases and changes sign, which implies amplification.  
We will discuss the problem in the hydrodynamic regime assuming that   electron-electron collisions dominate over other types of  scattering. (For discussion of crossover from hydrodynamic to ballistic  regime and  of the  Cherenkov instability in graphene see Refs.~\cite{Bylinkin2019Tight-BindingGraphene, Svintsov2019EmissionTransport}.)  
 The suggested mechanism  should work for any electronic spectrum, but calculations are  more compact and physically transparent  for parabolic spectrum characterized by the effective mass $m$. We thus restrict ourselves to the discussion of this case only. Qualitatively, the results are valid also for graphene, where role of $m$ is played by $E_F/v_F^2$. 
 
 The theoretical model that we develop in this work simplifies if, in addition to inequality \eqref{condition}, we also assume  that $s_1 \gg s_2$ [this inequality is different from Eq.~\eqref{condition} for $L_2 \gg L_1$].
This condition is not  strictly fulfilled (in our experimental situation   $s_1/s_2 \sim 2$   only) and  this could be one of the  possible  reasons  for  lack of quantitative agreement with experiment (see discussion and the supplementary materials). 

Since oscillations rapidly decay into passive regions, we consider dissipation in the active region only and put boundary conditions \eqref{BC} for all active strips. Dissipation per unit area is given by 
\begin{equation}\label{diss}
P=  
\left \langle N   \frac{ m \m v^2}{\tau } \right \rangle_{t,x},
\end{equation} 
where $\tau$ is the momentum relaxation time,  $N$ is local concentration  in the channel  and  $\m v$ is the   total  velocity   given by sum of the constant velocity related to \textit{dc} current and the oscillating component (see Ref.~\cite{PhysRevLett.114.246601}).  Equation \eqref{diss} implies averaging over active and passive regions, where $N$ and $\tau$ have different values.   
However, the main contribution comes from region 1, owing to the higher conductivity and resonant excitation. Hence, we skip the contribution of region 2 [here, in order to find dissipation averaged over the structure as a whole, the dissipation calculated in region 1 should be multiplied by the factor $L_1/(L_1+L_2)$].

Calculation of dissipation for  the  zero-current case is presented in Supplementary materials.   
We derived there an exact formula for dissipation, Eq.~(S21), and demonstrated that it  can be (with a good precision)  replaced by simplified equation obtained within so-called damped oscillator model [see Eq.~(S25) of Supplementary materials]. This, in turn, results in Eq.~\eqref{eq:2} for correction to the transmission coefficient.  
The latter equation perfectly fits  the observed  resonances provided that one replaces $\gamma_1$ with $\gamma_{\rm eff}.$  This gives additional confirmation of  the  plasmonic mechanism.              

Let us now assume that we apply driving \textit{dc} current in the system leading to  non-zero   
electron  drift velocity ${\cal V}_1$ in the active region.   In this case, 
the expression for dissipation in the active region should be  modified as follows:
\begin{align}
\nonumber
P_1 &=       \left \langle N_1(1+n_1) \frac{ m ({\cal V}_1+ v_1)^2}{\tau_1 } \right \rangle_{t,x}
\\
 & \approx \frac{m N_1}{\tau_1} \left( {\cal V}_1^2 +  \left \langle  v_1^2+ 2  {\cal V}_1  v_1 n_1\right \rangle_{t,x}\right)
\label{diss-J}
\end{align}
Here,  $v_1$ and $n_1=\delta N_1/N_1$ are  the radiation-induced corrections to  drift velocity  and normalized concentration, respectively.   Hence, radiation-induced correction to dissipation in  the active region   is given by
\be
\delta P_1=
 \frac{m  N_1}{\tau_1}   \left \langle  v_1^2+ { 2  {\cal V}_1  v_1 n_1}\right \rangle_{t,x}.
\label{deltaP}
\ee
The second term in this equation arises due to the presence of the current. Remarkably, this term depends on the phase shift between $n_1$ and $v_1$,  and  can be negative. This means that Eq.~\eqref{deltaP} is not positively defined and for special cases $\delta P$ could become negative.   This implies switching  from dissipation   to amplification.  
The full theoretical  description of the plasmonic crystal with two regions having arbitrary properties is quite cumbersome but 
can be essentially simplified for the case, when damping of plasmonic oscillations  in the passive region is sufficiently large, so that condition Eq.~\eqref{condition} holds.

We use the same approach as we used  to  study the response at zero current (see Supplementary material). Namely, we will describe the problem  phenomenologically, by using  the same  boundary conditions, Eq.~\eqref{BC}, as for the zero-current case. 

\begin{figure}[!ht]
\centering
\includegraphics[width=1.0 \linewidth, bb = 0 0 450 225]
{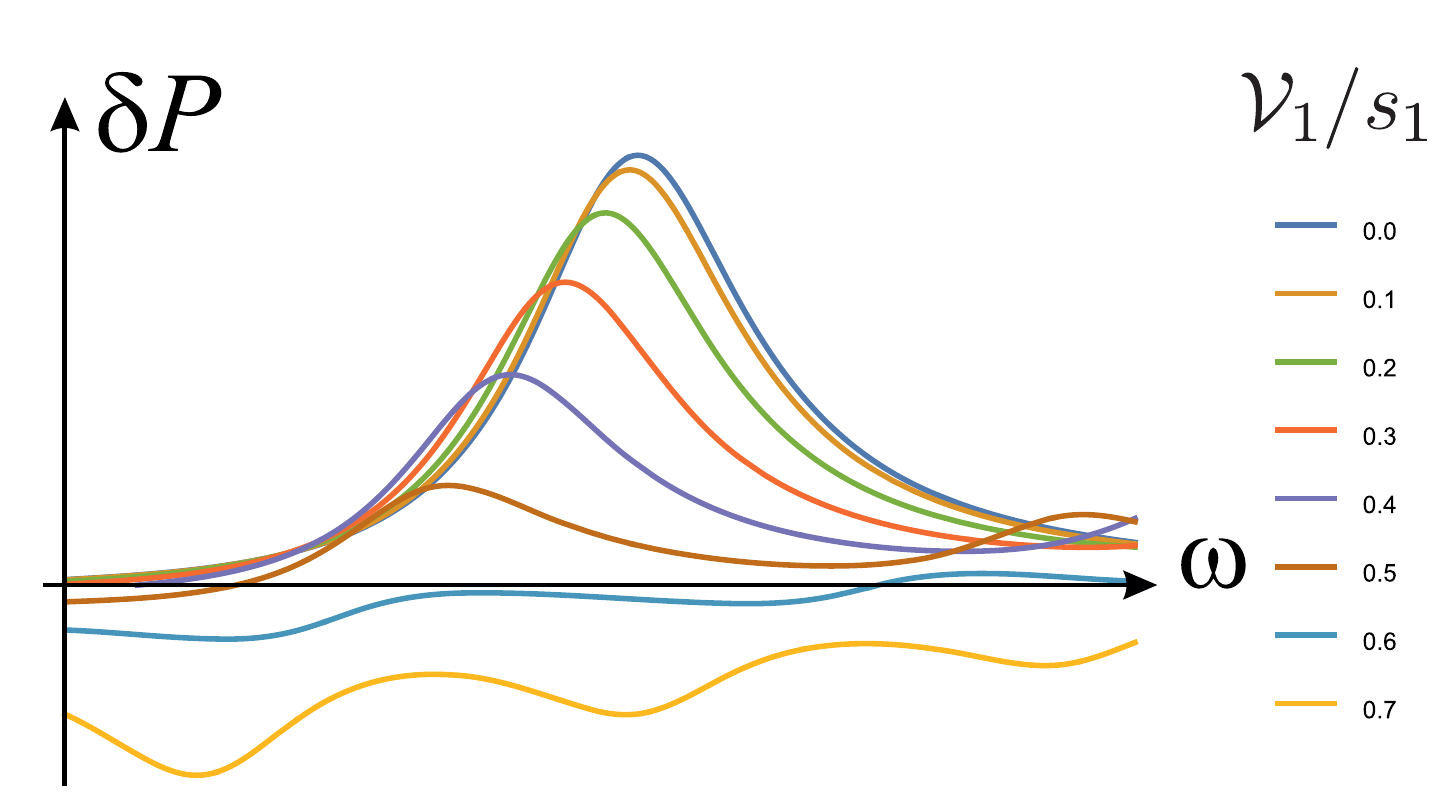}\par
\caption{Dissipation in the channel for different values of ${\cal V}_1/s_1$  }\label{fig2new}
\end{figure}

We  linearize  the hydrodynamic equations and search for solution in the following form $v_{1}=\delta v_1(x)e^{-i\omega t}~ +~ {\rm h.c.}$  and $n_1=\delta n_1(x)e^{-i\omega t} ~ +~ {\rm h.c.},$ where $\delta v_1(x)$ and $ \delta n_1(x)$
 obey
\begin{align}
& \left ({\cal V}_1 \frac{\p}{\p x}-i\omega+\gamma_1 \right) \delta v_1  +s_1^2 \frac{\p \delta n_1}{\p x} = \frac{F_0}{2 m},
\label{v1-current}
\\
& \left({\cal V}_1 \frac{\p}{\p x}-i\omega\right) \delta n_1   + \frac{\p \delta v_1}{\p x} =0.
\label{n1-current}
\end{align}
The oscillating correction to the current  is given  by $j_1=\delta j_1(x)e^{-i\omega t} ~ +~ {\rm h.c.},$
where 
\be
\delta j_1(x)=s_1^2\left[  {\cal V}_1 \delta n_1(x)+ \delta v_1(x) \right].
\ee
We solve  Eqs.~\eqref{v1-current} and ~\eqref{n1-current}  with boundary conditions  Eq.~\eqref{BC}, 
 substitute  $v_1$ and $n_1$ into Eq.~\eqref{deltaP} and average over time and space. 
 Thus obtained solution depends on the current, which is encoded in the drift velocity without radiation, ${\cal V}_1.$
Calculations are  very similar to the conventional calculation of the response of a single FET (see Refs.~\cite{Dyakonov1996DetectionFluid} and \cite{PhysRevB.73.125328}).  
The dependence of  total averaged
dissipation $\delta P =  [L_1/(L_1+L_2)] \delta P_1   $  on $\omega$ for different  ${\cal V}_1/s_1$ 
is shown  in Fig.~\ref{fig2new}.
 Qualitatively, this dependence  is similar to the experimentally observed one.  Indeed, as seen, with increasing the current, the  frequency  and the amplitude of the plasmonic resonance  decrease. For sufficiently large current,  $\delta P$ changes sign, which implies amplification. It is worth noting, however, that
 there appear some new peaks in the amplification, which are not seen in the experiment.
General analytical expression for dissipation simplifies to the form allowing an analytical solution if the quality factor of  resonance is high\sout{\red{,}}  so that its  frequency  is much larger than damping. For this case,  assuming  $L_1=L_2=L,$ we get
\be \label{dPres}
\delta P=\frac{ F_0^2 N_1  }{ m \pi^2}\frac{\gamma_{\rm eff}}{\left[\omega- \omega_1(x) \right]^2+\gamma_{\rm eff}^2/4^2} A(x),
\ee
where
\begin{align}\label{wx}
  \omega_1(x)&=\frac{\pi s_1}{L_1} (1-x^2),
  \\
  A(x) & =\frac{(1-3x^2)\cos^2(\pi x/2)}{(1-x^2)^2},
  \label{Ax}
\end{align}
is the frequency and amplitude of the resonance, respectively, which depend on $x={\cal V}_1/s_1.$
  \begin{figure}[!ht]
\centering
\includegraphics[width=0.9\linewidth,  bb = 0 0 441 312]
{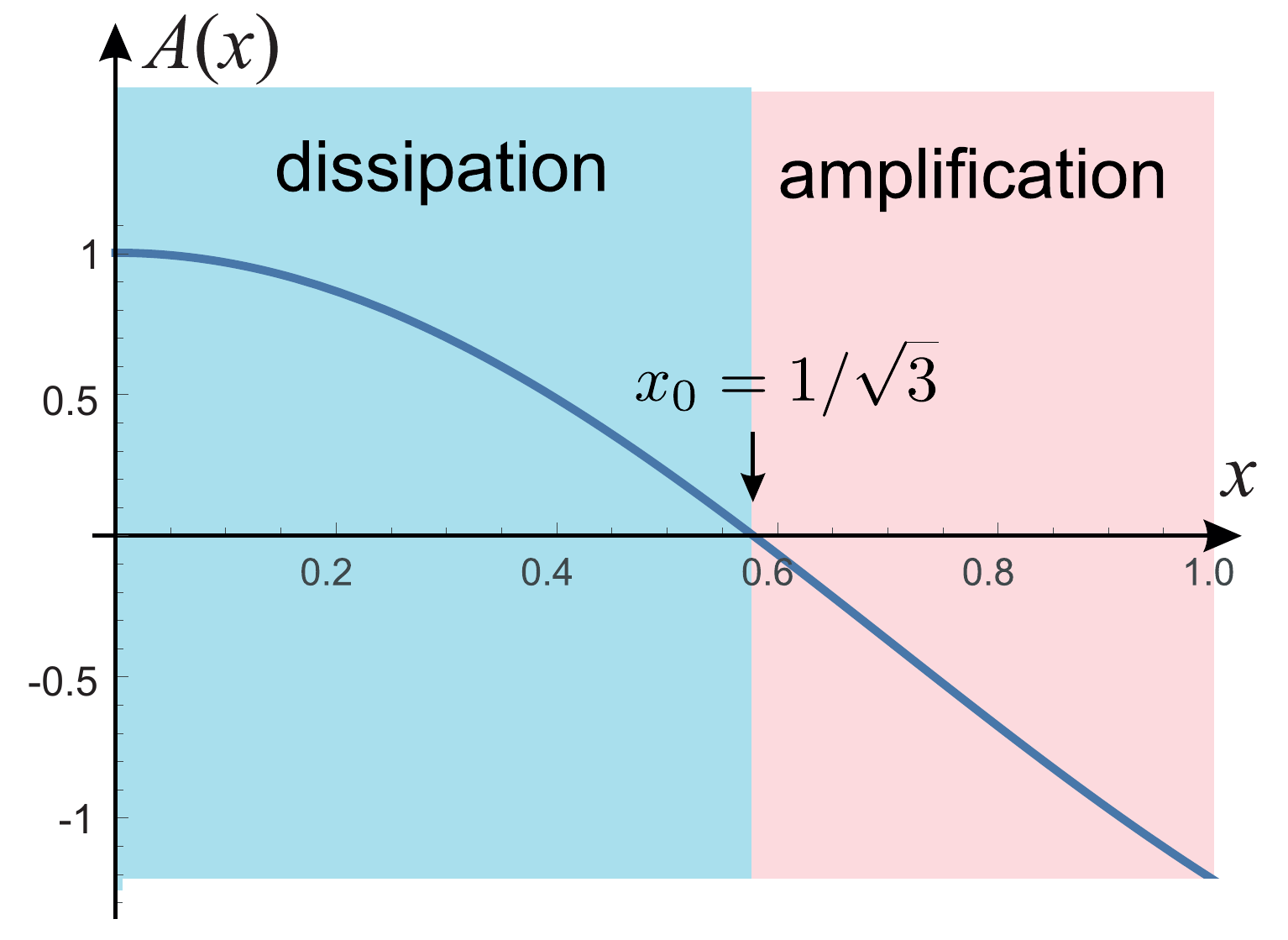}\par
\caption{The amplitude of the dissipation  as a function of $x={\cal V}_1/s_1.$  Dissipation-amplification transition corresponds to $x=x_0=1/\sqrt{3}. $ }\label{fig3new}
\end{figure}
Even if this analytical solution was obtained for conditions somehow far from the experimental ones, it captures main characteristics. It shows both the Doppler shift and the  possibility of  amplification. Indeed,  the frequency $\omega_1(x)$  depends on current   [see Eq.~\eqref{wx}], while     amplitude $A(x)$ turns to zero at certain value of  $x=x_0=1/\sqrt{3},$ which is the dissipation-amplification  transition point as shown in Fig.~\ref{fig3new}.

\section{Discussion} \label{sec:discussion}

Let us  briefly discuss other  known   mechanisms  of amplification and instability.     

First of all,  it is instructive to compare our theoretical model with the theory   developed in Ref.~ ~\cite{Mikhailov1998} for  grating-gated 2DEG in GaAs. This theory predicted that with the increase of the current (drift velocity) one may observe strong Doppler shift of plasma resonance followed by 100$\%$ transparency range and amplification. These predictions are very similar to our observations.
Moreover, in principle, Eq.~\eqref{fit_omega} can be obtained  from Eq.~(52) of \cite{Mikhailov1998} (see Supplementary material). However,  despite the good description by Eq.~\eqref{fit_omega}, the physical interpretation using the models of Ref.~\cite{Mikhailov1998}   is not possible. 
This model describes different physical situation, namely, in contrast to our system, the electron concentration is homogeneous in the absence of radiation.  

The second  reason, which is even more important,  is the fact, that to explain amplification within  this model one uses Cherenkov – like effects that require drift velocity in the active region to be higher than the plasma wave velocity.  Indeed,  it turns out that $  x={U_d}/{U_0}= {\cal V}_1/s_1, $ 
so that amplification is only possible for ${\cal V}_1>s_1.$ 
   At the same time, 
our experimentally determined plasma velocity ($s_1\approx 3.2 \times 10^8$ cm/s) is higher than the Fermi velocity in graphene. The latter represents the upper limit for drift velocity. Therefore, in our experiments  ${\cal V}_1 \ll s_1$.

In Ref.~\cite{Kachorovskii2012a}, the instability in gated plasmonic crystal was predicted  for the case when drift velocity exceeds plasma wave velocity in the passive region. This mechanism, however, was developed for ideal system with very low  momentum relaxation  rate  $\gamma_2 \ll s_2/L$ and cannot be applied for our system, where experimental conditions correspond to inequality \eqref{condition}.   Also, since experimental data show the strong evidence that the threshold value of velocity, corresponding to onset of amplification is smaller than the plasma wave velocity in the active region; we can exclude other mechanisms related to  Cherenkov-like plasmon instability  \cite{Matov1997a,Krasheninnikov1980a,Mikhailov1998,  Hwang2012FermiModification,  Aizin2016}.  

There are only a few theoretical predictions of plasma instability for drift velocities smaller than the plasma velocity. Most known is the instability based on amplified plasmon reflection at the cavity boundaries~\cite{Dyakonov1993a}. This instability is called Dyakonov-Shur instability. The evolution of \textit{dc} response with  the current increasing and approaching to instability threshold was described in Ref.~\cite{PhysRevB.73.125328}. It was predicted that current leads to decrease of the damping rate in the channel, which turns to zero on the instability threshold [see Eq.~(59) and (60) of Ref.~\cite{PhysRevB.73.125328}]. Evidently, such a scenario is not realized in our experiment, where the width of resonances does not essentially depend on the \textit{dc} current.

Although one cannot fully exclude the  mechanism based on so-called transit time instability  (it is  caused by exchange of excitation  between active regions due to carriers moving with saturated drift velocity in passive regions)   considered in Ref’s~\cite{Ryzhii2005, Ryzhii2006a, Koseki2016a},  there are some experimental  facts  against this mechanism. Most importantly, this mechanism implies a strong connection between  active regions via exchange of   excitation travelling through  passive regions. At the same time, the experimental data show that active regions behave like independent plasmonic resonators.

Importantly, in our work we do not suggest instability mechanism. Instead, we predict that current-driven amplification of the incoming radiation is possible without developing  any kind of plasma instabilities. 
Then, limitations ${\cal V}_1>s_1$ (needed for Cherenkov instability) and/or built-in asymmetry of the device (needed for well known Dyakonov-Shur instability \cite{Dyakonov1993a}) are lifted. We demonstrate theoretically that amplification is possible even if the drift velocity in the active region is smaller than the plasma wave velocity  (${\cal V}_1 < s_1$).


Let us now   discuss possible reasons, why the derived equations do not give  experimentally observed transparency gap. As we demonstrated,  in the absence of dissipation, for $s_1 \gg s_2,$ and  for zero current, dispersion equation \eqref{bands} reduces to the product of two sines, which give   resonance excitation frequencies. One can show that in the presence of dissipation and for nonzero current, the argument of sine corresponding to the active region transforms as follows
\be
\omega T_1 \rightarrow \frac{L_1 \sqrt{s_1^2 \omega(\omega+ i \gamma_1) -\gamma_1^2\mathcal V_1^2/4 }}{s_1^2- \mathcal V_1^2}. 
\ee
To find the  fundamental mode frequency, one should equal this expression to $\pi,$ which yields 
$\omega_1=\omega_1'+ i \omega_1''$, where  $ \omega_1''=-\gamma/2$ and   $(\omega_1')^2= (\pi s_1/L_1)^2 (1-x^2)^2- \gamma_1^2 (1-x^2)/4. $   
Hence,  $(\omega_1')^2$  turns to zero for $x=x_*,$ where 
\be 
x_*= \frac{\mathcal V_1^*}{s_1}=\sqrt{ 1- \left(\frac{\gamma_1 L_1}{ 2\pi s_1 }\right)^2} <1. 
\ee 
When $x$ exceeds $x_*,$ the solution for $\omega_1$   disappears because    $(\omega_1')^2$  can not be  negative.   
Therefore, $x_*$ corresponds to the onset of the transparency gap. 
Actually, this  gap does not have a sharp edge because $\omega_1'' \neq 0.$ 

The latter property  is similar to our  experimental observations. Indeed, Fig.~\ref{fit_omega} shows gap for  the center  of resonance peak. In fact, this peak is broadened due to the scattering 
within the  sufficiently large frequency interval about 1 THz (see Fig.~\ref{Fig8}, where 
smearing of resonances is shown in different shades of colors).  This gap does not show up in  our theory because we were able to perform calculations  only for the resonance case $\gamma_1  \ll \pi s_1/L_1. $  In this case,  $x_*  $ is larger than critical value of the dissipation-amplification transition, $x_0=1/\sqrt 3 $ (see Fig.~\ref{fig3new}.)               

We also note that our calculations were based on assumption $s_1\gg s_2, $ which is not well fulfilled in our experiment. Actually, in experiment, the  mismatch parameter $Z$ entering  Eq.~\eqref{bands} is not large (and, as one can show, depends on the current).  This 
 should  be taken into account when calculating the band structure of the plasmonic crystal and determining the transparency gap.
     
A more detailed analysis allowing to consider the case $s_1 \sim s_2$ 
     and take into account  that parameter  ${\gamma_1 L_1}/{ 2\pi s_1 }$ is not very small leads to very cumbersome equations that can be analyzed only  numerically. It is the subject of ongoing work and its results will be published elsewhere~\cite{Gorbenko2019}.

Finally, we note that interesting issue  for future study is the  response  to the radiation with  
parallel polarization. As we mentioned above, such a radiation does not lead 
to redistribution of charge even in the case of non-zero \textit{dc} current in perpendicular direction, so that plasmonic effects do not show up. What is non-trivial in this case is the role of  the viscosity of the electron liquid in the  channel~\cite{Torre2015PRB, Bandurin2016Science, Sulpizio2019}.  Indeed, because of  different  velocities in the passive and active strips (due to difference of scattering times), one can expect  viscosity-dependent friction between active and passive regions. This might lead to arising  of  Poiseuille-like  flows  oscillating with the radiation frequency~\cite{Sulpizio2019}.

\section{\label{sec:conclusion}Conclusion}

We have presented a study of gate and drain biasing on resonant 2D Dirac plasmons, excited by THz radiation, in grating gate graphene field effect transistor structures. These grating gate structures were used to create a periodic structure of highly-conducting “active” regions separated by low-conducting “passive” regions.

Without applied current, we observed the 2D plasmons with frequencies corresponding to the cavity modes  excited under  active grating-gate fingers. We have experimentally demonstrated that at the zero drain bias conditions, active regions act as independent ones.  Specifically, we found that  excited plasmons follow both the scaling laws of plasmon dispersion and the gate voltage dependencies predicted by standard physical models developed for a single active  cavity. 

The experimental results showed strong asymmetry of plasmonic resonances. The shape of such asymmetric resonances was very well fitted by the so-called damped oscillator simplified model, which, in turn, was in a very good agreement with the exact calculation of plasmonic dissipation peaks in the active regions. 

In the experiments with applied \textit{dc} current, we observed that increase of the current leads to a strong Doppler-like shift of the plasma resonances, followed by full transparency and THz light amplification phenomena. 

To interpret the experimental data, we have developed a phenomenological model of a plasmonic crystal formed by high and low plasma velocity regions (active and passive regions, respectively). The model captures basic physics of the problem and allowed qualitative  physical explanation of key experimental data.
In the case of the zero current it provided explanation why the multiple  high carrier density graphene cavities (model active regions) formed under gate fingers respond in the experiments with THz excitation as independent single cavity resonators with resonant frequency defined mainly by their dimensions. 
The main effect of the passive regions is  to provide  the plasmon leakage from the active regions, which explains  the  line widths of the observed plasmon resonances. 

For non-zero current, the phenomenological model describes basic tendencies together with qualitative description of physical mechanism of amplification taking place even for drift velocities smaller than the plasma velocities. The origin of the amplification phenomena is proposed as due to influence of \textit{dc} current on the phase shift between the carrier density and drift velocity induced by the THz radiation.

The presented model captures main trends and basic physics of the observed phenomena, but it does not provide quantitatively complete descriptions of all observed facts.It shows certain levels of discrepancies on the critical velocity and the threshold frequencies of the dissipation to amplification transition, predicting them much higher than the levels observed in the experiments.Also the model is not describing the transparency frequency intervals and do not demonstrate the  origin of the  interpolating Eq.~(\ref{fit_omega}), that provides an excellent fit of frequency versus drain voltage experimental dependencies for all cavities. Therefore, our results show a clear need for further, more advanced  theory of plasmonic crystals.

\hfill\break

 \textit{\textbf{Summarizing, we list our main results}}:
\begin{itemize}
\item	We have demonstrated current-driven excitation of Dirac plasmons in grating-gate graphene-channel transistor nanostructures leading to THz radiation amplification up to  room temperature. Specifically, we demonstrated that the plasmon resonances are red–shifted, undergo complete transparency to electromagnetic radiation followed by the amplification with more than 9 \% of gain and blue-shift of the resonant plasmon frequency in the THz range. 
\item	We have theoretically modeled the 2D Dirac plasmon system to explain the origin of the amplification phenomena – as due to influence of \textit{dc} current on the phase shift between the change of carrier density and drift velocity induced by the THz radiation. This radiation-induced correction to dissipation is sensitive to this phase shift and with increasing \textit{dc} current the dissipation becomes negative, which capture the basic physics behind  experimentally observed switching from dissipation to amplification.
\item	We have demonstrated, both theoretically and experimentally.that  transition to amplification regime occurs at drift velocities smaller than the plasma wave velocity. We theoretically found a critical value of drift velocity corresponding to the onset of amplification.
\item	We have discussed a leakage of plasma waves in the plasmonic crystals, which can limit plasmonic quality factor even in very clean structures. We derived analytical expression for leakage-dominated damping which shows that leakage can be controlled by gates covering passive regions of the plasmonic crystal.

Finally,  we would like to stress that all experimental results presented in this work were obtained at room temperature. We obtained amplification of the order of 10 percent that is far beyond the level of the fundamental limit for interband-transition-originated gain in monolayer graphene. Therefore, despite lacking a complete quantitative theoretical description, this work demonstrating room temperature THz amplification paves the way towards a  THz plasmonic technology  with a new generation of all-electronic, resonant, voltage- and current-tunable THz amplifiers.
\vspace{3.5mm}
\end{itemize}

\section*{\label{sec:acknowlegement} Acknowledgements}
We thank  V.~Ryzhii, S.~Mikhailov, D.~Svintsov, K.~Maussang, M.S.~Shur, M.~Dyakonov for many useful discussions.

The work was supported by JSPS KAKENHI (\#16H06361, \#16K14243, and \#18H05331), Japan; the  “International  Research Agendas” program of the Foundation for Polish Science co-financed by the European Union under the European Regional Development Fund for CENTERA (No. MAB/2018/9) and  by the Foundation for Polish Science through the TEAM project POIR.04.04.00-00-3D76/16 (TEAM/2016-3/25).
The work of V.Yu.K. was supported by Russian Foundation of Basic Research (Grant No. 20-02-00490) and by Foundation for the Advancement of Theoretical Physics and Mathematics “BASIS”. The work of V.V.P. was carried out within the framework of the state task.

%

\end{document}


\title{SUPPLEMENTAL MATERIAL
\\
for \\
Room Temperature  Amplification  of Terahertz Radiation by Grating-Gate Graphene Structures} 

\newcommand{\Sendai}{\affiliation{Research Institute of Electrical Communication, Tohoku University, Sendai 980-8577, Japan}}
\newcommand{\CENTERA}{\affiliation{CENTERA Laboratories, Institute of High Pressure Physics PAS, Warsaw 01-142, Poland}}
\newcommand{\CEZAMAT}{\affiliation{CEZAMAT Warsaw Technical University, Warsaw 02-346, Poland}}
\newcommand{\MLCC}{\affiliation{Laboratory Charles Coulomb, University of Montpellier and CNRS, Montpellier F-34095, France}}
\newcommand{\IRE}{\affiliation{Kotelnikov Institute of Radio Engineering and Electronics (Saratov Branch), RAS, Saratov 410019, Russia}}
\newcommand{\IOFFE}{\affiliation{ Ioffe  Institute, 194021 St. Petersburg, Russia}}

\author{Stephane Boubanga-Tombet}
\Sendai

\author{Wojciech Knap
\orcid{0000-0003-4537-8712}}
\Sendai
\CENTERA
\MLCC

\author{Deepika Yadav
\orcid{0000-0003-1240-9789}}
\Sendai

\author{Akira Satou
\orcid{0000-0002-4371-9344}}
\Sendai

\author{Dmytro~B.~But~\orcid{0000-0002-0735-4608}}
\CENTERA
\CEZAMAT

\author{Vyacheslav V. Popov}
\IRE

\author {Ilya V. Gorbenko}
\IOFFE

\author{Valentin Kachorovskii}
\CENTERA
\IOFFE

\author{Taiichi Otsuji
\orcid{0000-0002-0887-0479}}
\email{otsuji@riec.tohoku.ac.jp}
\Sendai

\date{\today}

\maketitle
\subsection{Transmission through array of conducting strips} \label{mihailov}

To analyze transmission through array of conducting strips, one can use theory developed in Ref.~\cite{Mikhailov1998} (see also Ref.~\cite{Ju2011}).
In the case when  parameter $2\pi \sigma_1/ c \sqrt{\epsilon}$ is small  ($2\pi \sigma_1/ c \sqrt{\epsilon} \ll 1$) the formulas for $T,R$ and $A$ (transmission, reflection, and absorption coefficients, respectively) can be obtained by expansion over this parameter.
 Indeed, as seen from Eq.~(35) of Ref.~\cite{Mikhailov1998} this parameter coincides with the ratio of the radiative decay $\Gamma$ to the momentum relaxation rate:
 \begin{equation}
 \frac{\Gamma}{\gamma_1} =  \frac{2\pi \sigma}{ c ~\sqrt{\epsilon} }.
 \end{equation}
 Since, we assumed  that inequality Eq.~(1) 
 is fulfilled, we find  $\Gamma \ll \gamma_1.$  Then, we find that expressions for  $T,R$ and $A$  [see  Eqs.~(40a), (40b) and (40c) of Ref.~\cite{Mikhailov1998}]  simplify.  In the vicinity of plasmonic resonance, we find
 \begin{align}
 &T= 1- \frac{\pi \sigma_1}{ c ~\sqrt{\epsilon} } \frac{\gamma_1^2}{\delta \omega^2 +\left(\frac{\gamma_1}{2}\right)^2},
\label{T}
 \\
 &R= \left(\frac{\pi \sigma_1}{ c~\sqrt{\epsilon}} \right)^2 \frac{\gamma_1^2}{\delta \omega^2 +\left(\frac{\gamma_1}{2}\right)^2},
  \label{R}
  \\
 & A=\frac{\pi \sigma_1}{ c ~\sqrt{\epsilon}} \frac{\gamma_1^2}{\delta \omega^2 +\left(\frac{\gamma_1}{2}\right)^2}.
 \label{A}
 \end{align}
Here $\delta \omega =\omega -\omega_1$ and $\omega_1$ is the frequency of the plasmonic resonance.
As seen from these equations,  \begin{equation} R \sim  \frac{\pi \sigma_1}{ c ~\sqrt{\epsilon} } A  \ll A \ll 1, \quad 1-T \approx A \ll 1,  \label{ineq} \end{equation}provided that condition (1) 
is fulfilled. This is in good agreement with  Fig.~4 of Ref.~\cite{Mikhailov1998}. Indeed, as seen from this figure, conducting strips, created by gate voltage,  change transmission coefficient of the system as a whole  by a  small correction (much smaller than unity). This implies that transmission through graphene strip is close to   unity, which is the case in our experiment.    This yields experimental proof that $2\pi \sigma_1/c \ll 1.$  Then, reflection is small, $\propto (\pi \sigma_1/\sqrt \epsilon c)^2 ,$ and can be neglected,  while $T~\approx 1-A$. Using  the Drude formula for conductivity $\sigma_1=e^2 N_1 /m\gamma_1,  $ one can rewrite   Eq.~\eqref{A} as follows
\be A=\frac{\pi e^2 N_1 }{ m c ~\sqrt{\epsilon}} \frac{\gamma_1}{\delta \omega^2 +\left({\gamma_1}/{2}\right)^2}.\ee
Comparing this equation with Eq.~\eqref{DO1}, we  find $A \approx {8 \pi P_{DO}}/{c E_0^2}  .$   It is convenient to rewrite this equation in a more general way   
\be
A =\frac{8 \pi \delta P}{c~\sqrt{\epsilon} E_0^2}, 
\label{A-final}
\ee
where $\delta P $ is radiation-induced correction to  the dissipation. Equations (3)
and \eqref{A-final} allow us to reduce the problem of calculation of transmission coefficient to the calculation of $\delta P.$   

\subsection{ Dissipation in the strip of width $L$}
As we discussed in the main text, active regions are disconnected at THz frequency  due to the inequality Eq.~(12)
The effect of passive regions can be taken into account by the following  replacing in  the final equations: $\gamma_1 \to \gamma_{\rm eff}.$ Hence, we can consider  a single active strip.      
The hydrodynamic equations for velocity $\m v$ and concentration $N$ in conventional 2D system with parabolic spectrum with mass $m$ read (for graphene away from the Dirac point, calculations presented below  should   also work if one   replaces $m$ with $E_F/v_F^2$):
\begin{eqnarray}
&& \frac{\p \mathbf v_1}{\p t} + (\mathbf v_1 \nabla) \mathbf v_1 +\gamma_1 \mathbf v_1 = -\frac{\p \Phi}{ \p \mathbf r } +\frac{F_0}{m} \cos (\omega t), 
\label{b1}
\\
&& \frac{\p N_1}{\p t} + {\rm div}(N_1 \mathbf v_1)=0 
\label{b2}.
\end{eqnarray}
Here $\gamma_1 =1/\tau_1$,   $\m F_0=e \m E_0,$  and $\m E_0$ is the electric field of the radiation,
\begin{equation}\Phi= \Phi(\m r,t)= \int d^2\m r^\prime K(\m r-\m r^\prime)\delta N_1(\m r ^\prime,t)  
\end{equation}
is the potential created by redistribution of the electron density, and $\delta N_1(\m r ,t)= N_1(\m r,t)-N_1$ is the deviation of the concentration from its equilibrium value.  Function $K(\m r) $ depends on the geometry of the problem. In particular, for gated 2D system,  $K(\m r) = (s_1^2/N_1) \delta (\m r)$,  where $N_1$ is the homogeneous  electron concentration in the absence of radiation and  $s_1$ is the plasma wave velocity in the active region.

\begin{figure}[ht!]
\centering
\includegraphics[width=0.55\linewidth, ,bb=0 0 557 708]{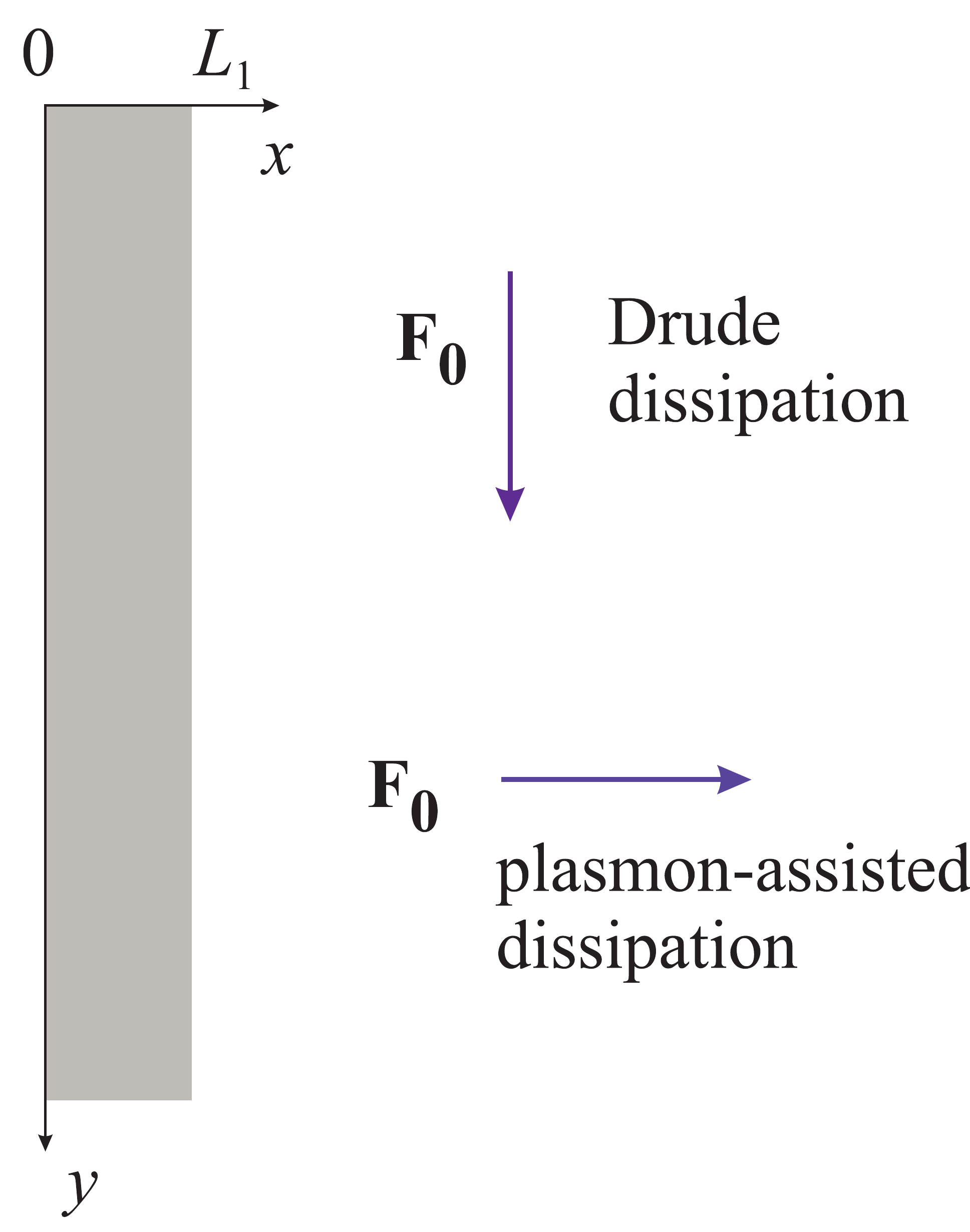}
 \caption{ Infinite strip of width $L_1$ in the external field $\m F_0.$   Dissipation depends on orientation of $ \m F_0.$       }
\label{fig1}
\end{figure}

Let us   introduce  a dimensionless  concentration
\begin{equation}
n_1=\frac{\delta N_1(\m r,t)}{N_1},
\end{equation}
 and linearize  Eqs.~\eqref{b1} and \eqref{b2}  with respect to  $n_1$  and  $\mathbf v_1.$ 
 Assuming  that  the system is gated, we get
\begin{eqnarray}
&& \frac{\p \mathbf v_1}{\p t}  +\gamma_1 \mathbf v_1 + s_1^2 \nabla n_1 =  \frac{\m F_0}{m}\cos (\omega t)
\label{l1}
\\
&& \frac{\p n_1}{\p t} + {\rm div} \mathbf v_1 =0,
\label{l2}
\end{eqnarray}
\\

\subsection{Drude dissipation}
Dissipation in the system depends on the orientation of the field.
For $\m F_0 \| \m e_y$,  both $n_1$ and $\m v_1$ do not depend on coordinates. Hence
$n=0,$ $v_x=0$
\begin{equation}
v_{1y}= \frac{ F_0}{2 m} \frac{1}{-i\omega +\gamma_1} e^{-i \omega t} + \quad h.c.
\label{vy}
\end{equation}
From Eqs.~(17)
and \eqref{vy}, we get the Drude dissipation:
per unit area:
\begin{equation}
{
P_{\rm Dr}=\frac{ N_1   F_0^2}{2 m} \frac{\gamma_1}{\omega^2+ \gamma_1^2}
}
\label{Drude}
\end{equation}
\subsection{Plasmon-assisted dissipation}
For $\m F_0 \| \m e_x,$ external field excites plasmons in the strip. In this case, $v_y=0$ and $v_x$ and $n$ can be written as
\begin{align}
&v_{1x}=\delta v_1(x)e^{-i\omega t}+ \delta v_1^*(x) e^{ i\omega t},
\\
&n_1=\delta n_1(x)e^{-i\omega t}+ \delta n_1^*(x)e^{ i\omega t}.
\end{align}
The amplitudes $\delta v_1$ and $ \delta n_1$ obey
\begin{align}
& (-i\omega+\gamma_1) \delta v_1  +s_1^2 \frac{\p \delta n_1}{\p x} = \frac{F_0}{2 m},
\label{v1}
\\
& -i\omega \delta n_1   + \frac{\p \delta v_1}{\p x} =0.
\label{n1}
\end{align}
These equations should be solved with the boundary conditions $\delta v_1  (x=0)= \delta v_1 (x=L_1)=0$  (absence of current at the edges of the system). Direct calculation yields for the  velocity amplitude
\begin{equation}
\delta v_1 = \frac{F_0}{ m(\gamma_1 - i\omega) } \frac{ \sin(kx/2)\sin[k(x-L_1)/2]}{\cos(kL_1/2)},
\end{equation}
where
\begin{equation}
k=\frac{\sqrt{\omega(\omega+ i\gamma_1)}}{s_1}
\label{k}
\end{equation}
Now, dissipation is inhomogeneous, so that we need to average both over $t$ and over $x$. The averaged  plasmon-assisted dissipation per unit area looks (we skip contribution of passive regions)
\begin{widetext}
\begin{align}
&P_{\rm pl}=N_1 L_1 \left \langle \frac{m v_1^2}{\tau_1} \right \rangle_{x,t} = 2N_1 m \gamma_1  \int\limits _0^{L_1} 
\frac{dx}{L_1}~ |\delta v_1(x)|^2  \nonumber \\
&
  = \frac{2  N_1 F_0^2}{m} \frac{\gamma_1}{\omega^2 +\gamma_1^2}  \frac{1} {\cos (k L_1/2) \cos (k^* L_1/2)} \int\limits _0^{L_{1}} \frac{dx}{L_1} \sin(kx/2)\sin[k(x-L_1)/2]\sin(k^*x/2)\sin[k^*(x-L_1)/2].
\label{Ppl}
\end{align}
\end{widetext}
The plasmonic resonances are encoded in the factor $\cos (k L_1/2) \cos (k^* L_1/2)$ which turns to zero (for $\gamma_1=0$) at the resonant plasmonic frequencies
\begin{equation}
\omega_{1}(N)= \omega_1(1+2N),\quad N=0,1,2,\ldots
\end{equation}
where $$\omega_1=\frac{\pi s_1}{L_1} $$
is the fundamental plasmonic frequency in the active region. 
It is worth noting that homogeneous external field excites only  odd modes with $1+2N,$ while even modes $2N$ remain silent although they are present in the full set of resonant excitations.
Although Eq.~\eqref{Ppl} looks quite complicated, it dramatically simplifies in the resonant regime: $$\delta \omega=\omega- \omega_1(N)  \ll \omega , \quad \omega_1 \gg \gamma_1.$$
In this case,
for $N=0$ (fundamental plasma mode),
after some algebra,  we find
\begin{equation}{
P_{\rm pl}=\frac{N_1 F_0^2}{ \pi^2 m  }  \frac{\gamma_1}{(\delta \omega)^2 +  (\gamma_1/2)^2 }
}
\label{plasmon}
\end{equation}
 Comparing the amplitudes of the Drude and plasmon-assisted dissipation, we get
\begin{equation}{
 \frac{P_{\rm Dr} (\omega=0)}{P_{\rm pl}(\delta \omega =0)}= \frac{\pi^2}{8} \approx  1.23.
  }
  \label{ratio}\end{equation}

\begin{figure}[ht!]
\centering
\includegraphics[width=0.75\linewidth]{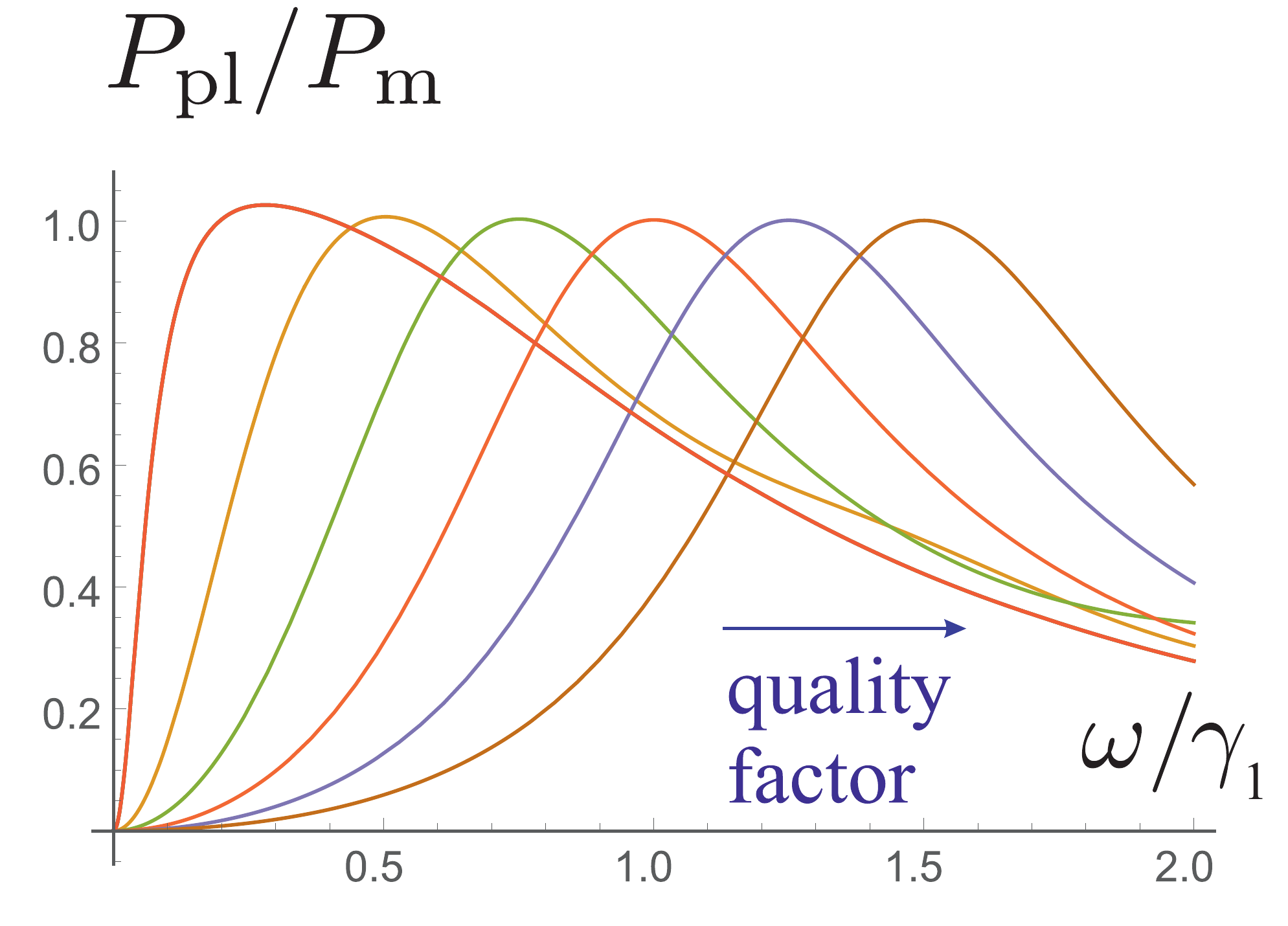}
 \caption{ Dissipation calculated by the use of Eqs.~\eqref{Ppl} and \eqref{Pm-supp} for different values of $\omega_1/\gamma_1=0.25,~0.5,~0.75,~1.0,~1.25,~1.5$ increasing  from  left  to right.          }
\label{fig2}
\end{figure}

Let us compare now Eqs.~\eqref{Drude}, \eqref{plasmon},   and \eqref{ratio} with Fig.~5 of Ref.~\cite{Mikhailov1998}.   First of all, from Fig.~5 of Ref.~\cite{Mikhailov1998} we see that amplitude of the Drude peak coincides with the amplitude of dissipation in the plasmonic resonance. This observation is in a reasonably good agreement with Eq.~\eqref{ratio}.  Second observation related to the widths of the Drude and plasmonic  peaks. As seen from  Fig.~5 of Ref.~\cite{Mikhailov1998} the width of the Drude peak is twice larger than the width of the plasmonic resonance. This observation is in excellent agreement with  Eqs.~\eqref{Drude}, \eqref{plasmon}. Hence, we come to the conclusion that  peaks in the transmission coefficient observed in   Ref.~\cite{Mikhailov1998} for different polarizations of the radiation can be well interpreted as the Drude (parallel polarization) and plasmonic (perpendicular) peaks. A more detailed analysis of this statement is presented in the next subsection.
As one can see from   Fig.~5 of Ref.~\cite{Mikhailov1998}, for small resonance frequencies, when quality factor of plasmonic resonance is smaller than unity, the peaks are asymmetrical functions  of frequency in contrast to  symmetrical Lorentz peak given by Eq.~\eqref{plasmon}.   To describe dissipation for small quality factors one can use exact equation Eq.~\eqref{Ppl}.    The result of calculations are shown in Fig.~\ref{fig2}. As seen,  asymmetry increases with decreasing the quality factor $\omega_1/\gamma_1$ in a very good agreement with experimental data. All curves are normalized to the value
\be P_{\rm m}= \frac{4 N_1F_0^2}{\pi^2 m \gamma_1},  \label{Pm-supp} \ee
which gives the maximal value of the dissipation for the case of high-quality factor [see Eqs.~\eqref{plasmon}].

One can compare results obtained with the use of the exact formula Eq.~\eqref{Ppl} with the results obtained within  damped oscillator model. Dissipation in this model  is given by  the  Drude  dissipation formula with the  plasmonic-induced frequency shift
\begin{equation}
P_{\rm DO}=\frac{N_1 F_0^2}{2 m} \frac{\gamma_1}{\left(\omega - \frac{\omega_1^2}{\omega}\right)^2+ \gamma_1^2}.
\label{DO}
\end{equation}
For the case of high-quality factors, $\omega_0 \gg \gamma,$  this equation simplifies even more
\begin{equation}
P_{\rm DO}\approx\frac{N_0 F_0^2}{8 m} \frac{\gamma_1}{\left(\delta \omega \right)^2+ \gamma_1^2/4},
\label{DO1}
\end{equation}
where $\delta \omega =\omega- \omega_1.$  Although resonance equation \eqref{DO1} is symmetric function of $\delta \omega,$ the general equation \eqref{DO} shows a strongly asymmetrical peak.     
 Equation \eqref{DO} describes very well  the shape of the resonance at the fundamental frequency, as illustrated in Fig.~3   for the same values of quality factors as in Fig.~\ref{fig2}
 \begin{figure}[ht!]
\centering
\includegraphics[width=0.4\textwidth]{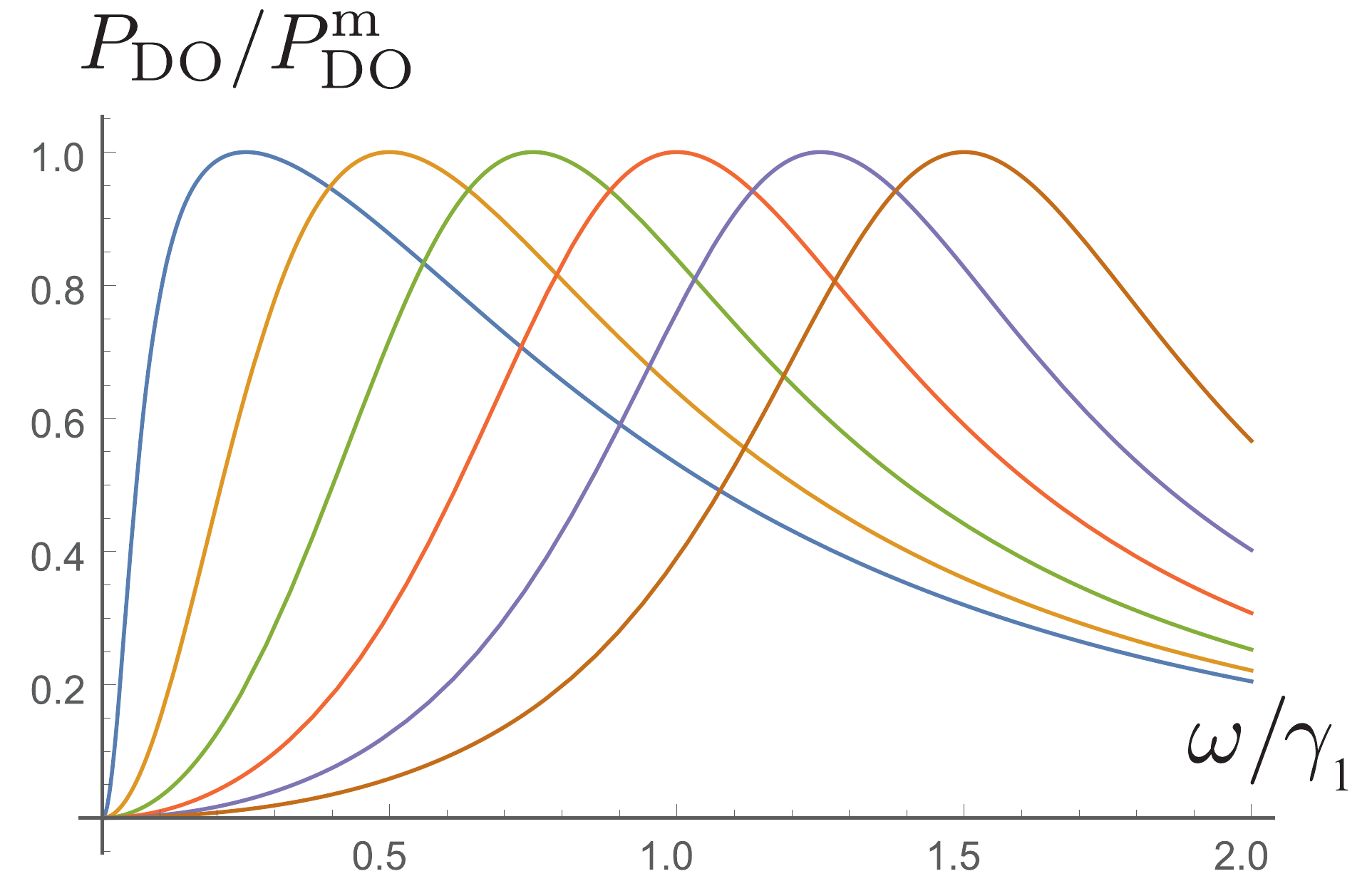}
 \caption{ Dissipation calculated by the use of Eq.~\eqref{DO} for different values of $\omega_1/\gamma_1=0.25,~0.5,~0.75,~1.0,~1.25,~1.5$ increasing  from  left  to right.          }
\label{fig3}
\end{figure}
  There are however some differences between exact formula \eqref{Ppl} and Eq.~\eqref{DO}. First of all, the amplitude  of  the dissipation is slightly different: $P_{\rm DO}^{\rm m}/P_m= \pi^2/8.$  Second, the shape of the curves also slightly different, and the difference increases with decreasing  the quality factor. We illustrate last statement in Fig.~\ref{fig4}, where we compare relative dissipation obtained from two equations for low-quality factor $\omega_1/\gamma_!=0.5.$      We also note that the damped plasmon model does not allow us to calculate the amplitude of plasmonic mode  with $N \neq 0.$ By using Eq.~\eqref{Ppl}, we find  in the resonance approximation   close to $N-$th resonance
   \begin{equation}
P_{\rm pl}^N=\frac{N_1F_0^2}{ \pi^2 m  (1+2N)^2}  \frac{\gamma_1}{(\delta \omega)^2 +  (\gamma_1)^2/4 },
\label{plasmonN}
\end{equation}
where $\delta \omega =\omega-\omega_N.$
Hence,
\begin{equation}
\frac{P_{\rm pl}^N(\delta\omega=0)}{P_{\rm pl}^{N=0}(\delta\omega=0)}=\frac{1}{(1+2N)^2}
\label{PN}
\end{equation}

\begin{figure}[!ht]
\centering
\includegraphics[width=0.85\linewidth]{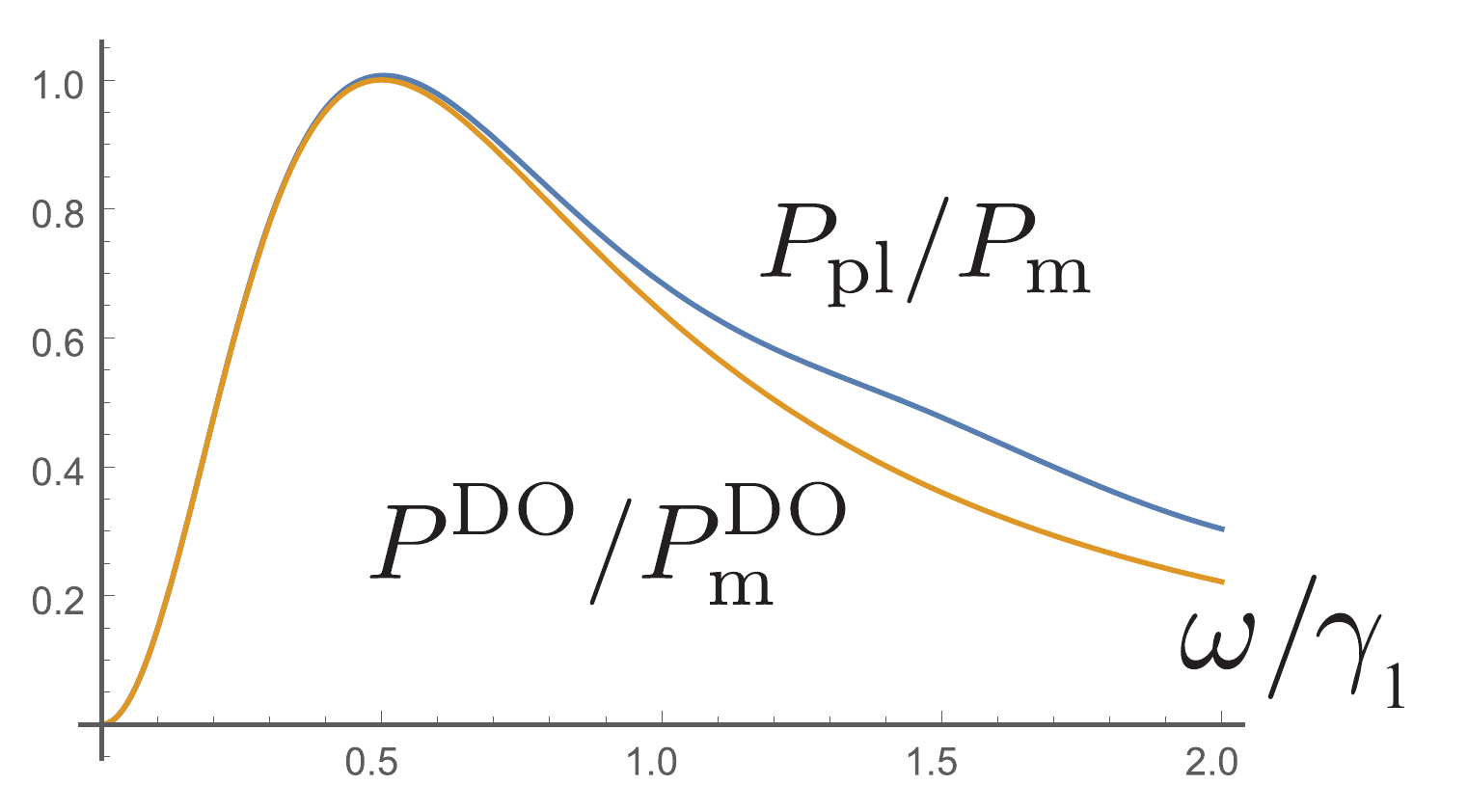}
 \caption{ Comparison of resonance shape found from  Eq.~\eqref{Ppl} and  Eq.~\eqref{DO}    for $\omega_1/\gamma_1=0.5$.          }
\label{fig4}
\end{figure}

\begin{figure}[!ht]
\centering
\includegraphics[width=0.85\linewidth]{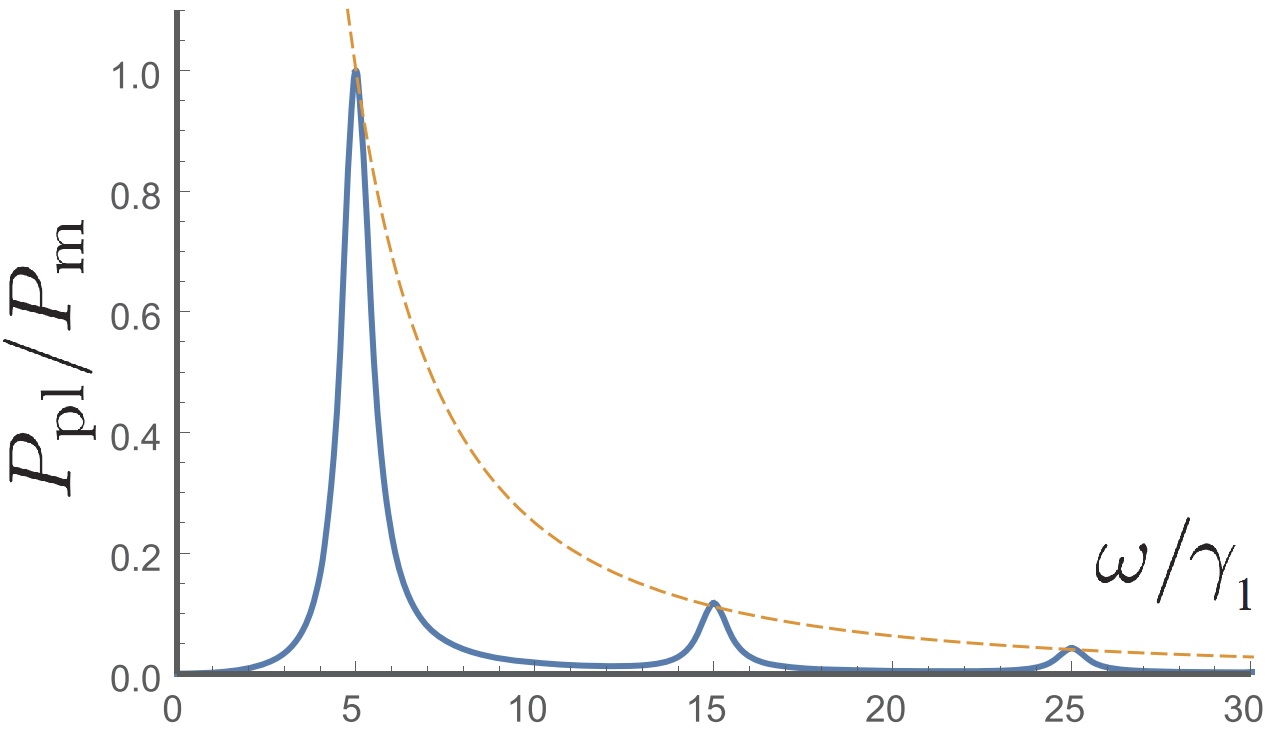}
 \caption{ Several resonances for $\omega_1/\gamma_1=5$.  Dashed line corresponds to Eq.~\eqref{PN} }
\label{fig5}
\end{figure}

Before closing this section, we note that Eq.~\eqref{Ppl} describes all resonances. In Fig.~\ref{fig5} we illustrate it for $\omega_1/\gamma_1=5$.

\subsection{Leakage  of plasmons from active to passive region}
Above, we assumed that plasmons are localized in the active regions.
Let us now  briefly discuss  the leakage of plasmonic excitation, which arise due to finite coupling with the passive regions (such leakage was previously predicted for a single gate problem 
in Ref.~\cite{Popov2018}). 
We  assume that damping rates in the active and passive regions are different, $\gamma_1  < \gamma_2$, which is the case in our experiment.
We also assume that $\gamma_2 L_2 \gg s_2,$ which means that different active regions are effectively  
disconnected so that without loss of generality, we can put $L_2=\infty$.
Then, we have three regions:
active region, $|x|<L_1/2$ and two passive regions: $-\infty<x<-L_1/2$ and $L_1/2<x<\infty.$  
We use the following boundary conditions (BC) between regions, $s_1^2v_1=s_2^2v_2,~s_1^2n_1=s_2^2n_2$ (for $x=\pm L_1/2$) which represent current and energy conservation at the boundary (see more detailed discussion in Ref.~\cite{Kachorovskii2012a}).  Plasmonic oscillations in the passive region can be described by the same equations as  Eq.~\eqref{v1} and \eqref{n1} with the obvious change of parameters. In the active region there are two solutions with wave vectors $\pm\sqrt{\omega(\omega+i\gamma_1)}/s_1,$ while in the passive regions,  
there is a solution with the 
wave vector $\sqrt{\omega(\omega+i\gamma_2)}/s_2,$  for $x>L_1/2$ and  solution with wave vector   $-\sqrt{\omega(\omega+i\gamma_2)}/s_2,$ for $x<L_1/2$. Hence, we have four independent solutions and, therefore, four independent  coefficients, which can be written as a vector  $\m a= (a_1,a_2,a_3,a_4)$.  This vector  should  be found from four BC.  We skip this quite standard calculation  and limit ourselves to the discussion of general idea and presenting the final result.  Using  BC one can  write equation for $\m a$ in the matrix form: $\hat A ~\m a =\m c,$ where  vector $\m c \propto E_0$ describes coupling to external field. Writing solution of this equation as $\m a = \hat A^{-1} \m c,$ we find that resonances in our system corresponds to poles of function $1/ {\rm det} \hat A$. Standard calculation yields (we skip all irrelevant coefficients)
${{\rm det} \hat A} \propto { \exp\left[{2i L_1   \sqrt{\omega (\omega + i \gamma_1)}}/{s_1} \right] - (u_1+u_2)^2/(u_1-u_2)^2}$,
where $u_{1,2} =s_{1,2}/\sqrt{1+ i \gamma_{1,2}/\omega}.$
Assuming that $\gamma_{1,2} \ll \omega$ and expanding this equation over $\delta\omega=\omega-\omega_1$ 
 we find expressions for effective damping and resonance frequency correction:  
\begin{align}
& \gamma_{\rm eff}= \gamma_{1}+\frac{2s_1}{L_1} \ln\left( \frac{s_1+s_2}{s_1-s_2}\right),
\label{leakage_1}
\\
& \omega_1^{\rm eff}= \omega_1 - \frac{\pi (\gamma_2-\gamma_1) s_1 s_2}{(s_1^2-s_2^2)\left[\pi^2+ \ln^2(\frac{s_1+s_2}{s_1-s_2}). \right]} ,
\label{real-part_1}
\end{align}
The effective width of the plasma cavities $L_1$ should include also correction related to the fringing effect, that leads to the following correction of the plasmon frequency: 

\begin{align}
\delta \omega_{1}^{\rm fr} = \frac{2d}{L_{1}}\omega_{1},
\label{d_omega_2}
\end{align}  

Second term in the r.h.s. of  Eq.~\eqref{leakage_1} accounts for  damping  of oscillation caused by leakage effect, while the second terms in the r.h.s. of  Eq.~\eqref{real-part_1}  yields small correction to resonance frequency caused by coupling between active and passive regions. (To avoid confusion, we notice that Eqs.~\eqref{leakage_1} and \eqref{real-part_1} are not valid for $s_2 \to s_1.$ First of all, we assumed that   ($\omega_1^{\rm eff} - \omega_1)/\omega_1 \ll 1.$ Also, at $s_2\to s_1$ the width of the resonance becomes much larger than $\omega_1,$ so that the  resonance disappears). 

Expressions for  $\gamma^{\rm eff}$ and $ \omega_1^{\rm eff}$ essentially simplify for $s_2 \ll s_1$:
\begin{align}
&  \gamma^{\rm eff}= \gamma_1+\frac{4s_2}{L_1}
\label{leakage}
,\\
&\omega_1^{\rm eff} = \omega_1 -\frac{ (\gamma_2-\gamma_1)  s_2}{\pi s_1 } ,
\label{real-part}
\end{align} 

We notice that leakage-induced contribution to the damping is large, which explains why width of resonance is much larger than $\gamma_1$ extracted from the transport measurements. On the other hand, negative correction to the real part of the frequency  is small. However, it can be responsible for small deviation of the experimentally observed resonance frequency from the theoretical value obtained within the model of an  isolated  active strip. 

\begin{figure}[!hb]
\centering
(a)\includegraphics[width=0.94\linewidth]{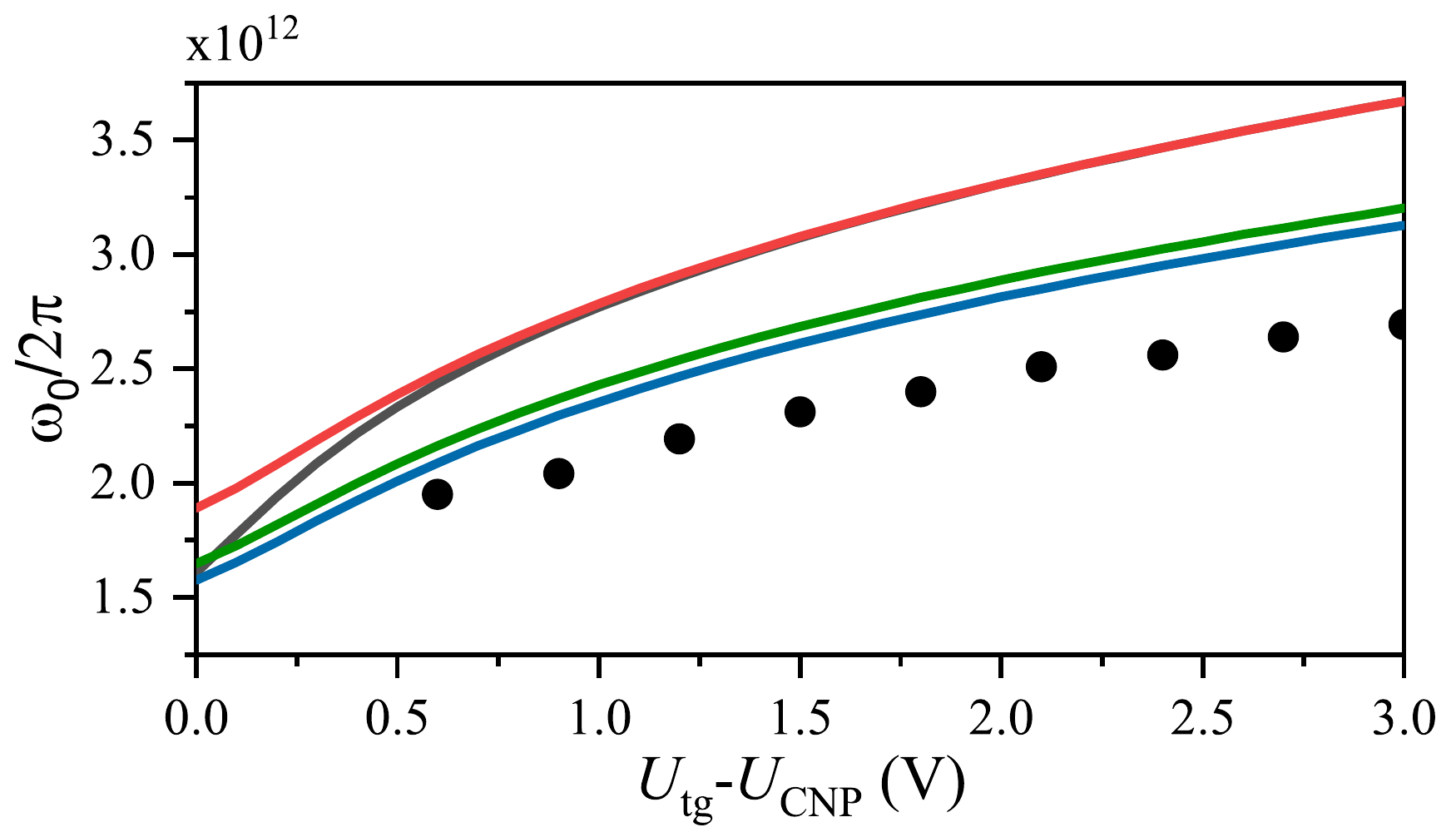}
(b)\includegraphics[width=0.94\linewidth]{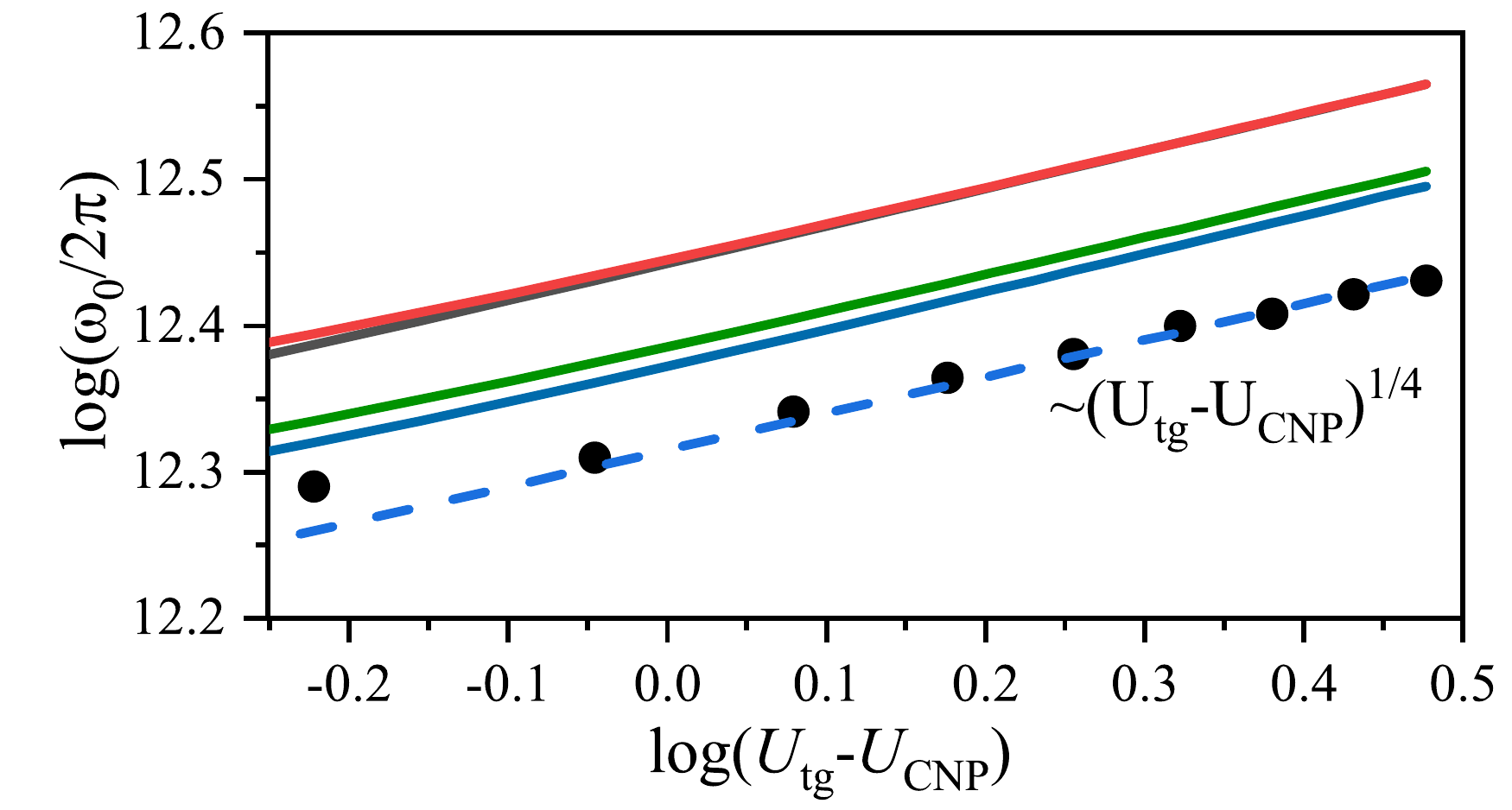}
\caption{
a) Plasma frequencies as a function of the gate voltages for the cavity C1. Dots are experimental points. The black line represents calculations according to Eq.~(8) 
and the red line is Eq.~(7).
The blue line is calculated according to Eq.~(7)
with additional correction due fringing effect Eq.~\eqref{d_omega_2}, and the green line is calculated taking into account additionally plasmon leakage phenomena as described by Eq.~\eqref{real-part}. b) The same results as in a) but in log-log scale. The dotted line shows $\nicefrac{1}{4}$~power dependence typical to plasmons in graphene. }
\label{Fig_FREQ_CALC}
\end{figure}

Numerical calculations of the plasma frequency versus a gate voltage for the cavity C1 are shown in Fig.~\ref{Fig_FREQ_CALC}(a). One can see that calculated values are slightly higher than experimental ones. The black line represents calculations according the Eq.~(8)
and the red line is Eq.~(7).
One can see that more advanced description (Eq.~(7))
gives correction to plasma frequency only in the region of small carrier densities where Fermi level is comparable to the thermal energy $kT$. The blue line is calculated according to Eq.~(7)
with additional correction due to fringing effect approximated by Eq.~\eqref{d_omega_2}. Finally, the green line is calculated taking into account additionally plasmon leakage phenomena as described in by Eq.~\eqref{real-part}. One can see that the plasmon leakage correction to the resonant plasma frequency is relatively small. In Fig.~\ref{Fig_FREQ_CALC}(b) the same results are shown in log-log scale demonstrating typical for graphene $\nicefrac{1}{4}$~power dependence. 
One can see that the theoretical calculations made without any fitting parameters reproduce relatively well the absolute values and the functional frequency versus gate voltage dependency. As already mentioned in the main text, the small discrepancies may result from the uncertainty of h-BN layer thickness determination (precision  10\%) and unknown exact value of the dielectric function. In fact, in the literature, one can find this value in the range from 3.5 to 5~\cite{Laturia2018DielectricBulk,Yu2013InteractionCapacitance}. 
It should be stressed however, that independently of small numerical discrepancies, the observed plasma frequency follows closely $\nicefrac{1}{4}$~power dependence on gate voltage (carrier density) typical for plasmons in graphene~\cite{Ryzhii2007PlasmaHeterostructures} - as shown by dotted line in Fig.~\ref{Fig_FREQ_CALC}(b).  

\subsection{Frequency of plasmonic excitations in the grating-gate structure}
In this subsection, we demonstrate how to derive Eq.~(10)
from Eq.(52) of Ref.~\cite{Mikhailov1998}. To this end, one should first determine $\omega_1$ from  this equation (by putting  drift velocity in the electron liquid to zero) and then rewrite this equation in terms of $\omega_1$:
\begin{widetext}
\be
\frac{\omega_1^2(x)}{\omega_1^2}=\frac{1- \Delta/2}{1-\Delta} + x^2- \sqrt{ \frac{\Delta^2}{4(1-\Delta)^2}  + 4 x^2 \frac{1-\Delta/2}{1-\Delta} },
\label{52-of-Mihailov}
\ee 
\end{widetext}
where $\omega_1(x)$ is the resonance frequency of the plasmon oscillations  as a function of  
and
$x=  q {\cal V}_1 /\omega_1,   $   
where  $q$ is the plasmon wave-vector.  Since  we demonstrated experimentally that $\omega_1= s_1 q,$ we find
$x= {\cal V}_1/s_1.$
Equation Eq.~\eqref{52-of-Mihailov} transforms into Eq.~(10)
if we take 
\be
\Delta=\frac{2\zeta}{1+ 2\zeta}.
\ee
Assuming that ${\cal V}_1 \propto U_d$ we arrive at  Eq.~(10).
It is worth stressing again that the  onset of the transparency  band corresponds to ${\cal V}_1=s_1,$ so that amplification starts for ${\cal V}_1>s_1.$    

%